\documentclass[aps,11pt]{revtex4}
\usepackage{amsmath}
\usepackage{amssymb}
\usepackage{graphicx}

\newcommand\revise[1]{#1}
\newcommand\new[1]{#1}

\newcommand{\Section}[1]{\section{#1} \setcounter{subsection}{0}}

\begin{document}
\title{Geometry of physical dispersion relations}
\author{Dennis R\"atzel}
\author{Sergio Rivera}
\author{Frederic P. Schuller}
\address{Albert Einstein Institute\\ Max Planck Institute for Gravitational Physics\\ Am M\"uhlenberg 1, 14476 Golm, Germany}
%\address{Institut f\"ur Astronomie und Astrophysik, Universit\"at Potsdam, 14476 Potsdam, Germany}
%\address{Max Planck Institute for Gravitational Physics\\ Albert Einstein Institute\\ Am M\"uhlenberg 1, 14476 Golm, Germany}
\begin{abstract}
To serve as a dispersion relation, a cotangent bundle function must satisfy three simple algebraic properties. These conditions are derived from the inescapable physical requirements to have predictive matter field dynamics and an observer-independent notion of positive energy. Possible modifications of the standard relativistic dispersion relation are thereby severely restricted. 
\revise{For instance, the dispersion relations associated with popular deformations of Maxwell theory by Gambini-Pullin or Myers-Pospelov are not admissible.}
%The remaining ones provide the 
%physically appropriate Finslerian extension of Lorentzian geometry. 

\end{abstract}
\maketitle

%\newpage
\section*{INTRODUCTION}\enlargethispage{2cm}
In the standard description of relativistic matter, the dispersion relation that governs the behaviour of matter is given in terms of a Lorentzian metric. In recent years, however, numerous authors have made a case for various modified dispersion relations \cite{ling2006,lopez2009,barcelo2007,rinaldi2007,garattini2010,bazo2009,gregg2009}, mostly motivated from specific approaches to quantum gravity \cite{kostelecky1989,gambini1998,alfaro1999,myers2003,alfaro2002,latorre1994,magueijo2002,gibbons2007,lukierski1995,glikman2001,girelli2007} or other particular physical and mathematical models \cite{sindoni2008,sindoni2009,sindoni2009-2,chang2008,laemmerzahl2009,rubilar2002,hehl2003,laemmerzahl2005,drummond1980,liberati2001,jacobson2006,tavakol2009,chang2008-2,huang2007,duval2008,wana2009,mignemi2007,Panahi2003,perlick2006,kouretsis2010,Garasko2007}. 

In this paper, we address the question of what can be said about {\it dispersion relations in\,general}, and most importantly how they are restricted in principle, independent of their physical motivation. 
Starting only from the fundamental physical requirements that the underlying theory be predictive and that there be a well-defined notion of positive energy, we show that there are three inescapable algebraic conditions that a modified dispersion relation in classical physics must satisfy. This severely restricts possible modifications to the standard relativistic dispersion relation.
More precisely, the central result we arrive at in this article is that
 \begin{center}{\it  A dispersion relation must be given by a cotangent bundle function $P$ that is\\
 a reduced, bi-hyperbolic and energy distinguishing homogeneous polynomial in each fibre.} \end{center}
These conditions, whose technical definition and justification will be the subject matter of the first four sections of this article, all essentially root in the application of well-known results from the theory of partial differential equations, real algebraic geometry and convex analysis to the kind of questions one considers in classical physics.
\newpage
Remarkably, the above conditions are not only physically necessary, but also mathematically sufficient to set up the entire kinematical machinery one needs in order to give physical meaning to quantities on spacetimes with such dispersion relations. The central r\^ole is played here by Gauss and Legendre maps, which provide a proper duality theory between cotangent and tangent spaces. Their existence is far from trivial once one leaves the Lorentzian metric framework, and indeed we will see that one requires some rather sophisticated real algebraic geometry that was originally conceived in the solution of Hibert's seventeenth problem. Once these theoretical issues are clarified, though, one sees that indeed everything that one wants from a Lorentzian metric in standard relativity is equally well afforded by a cotangent bundle function with the said three properties. And, this is maybe the most important point, only by such. Thus the present work defines the outer boundaries of what constitute physically viable classical spacetime structures.\\ 

To arrive at the above results requires numerous steps and occasional asides on mathematical techniques. The generality of the obtained result, however, makes this worth the effort. In order to provide the reader with an intuition for how the argument proceeds, we now briefly outline what is shown in each section.

In section \ref{sec_covdisprelmassless}, we study how linear matter field dynamics on an arbitrary tensorial geometry give rise to a massless dispersion relation. This will clarify two important points. On the one hand, we will see that any massless dispersion relation must be described by a cotangent bundle function $P$ that induces a reduced homogeneous polynomial in each cotangent space. This is the first of the three algebraic conditions identified in this work. On the other hand, it reveals the close link between dispersion relations and the underlying spacetime geometry seen by specific fields. Thus the restrictions on dispersion relations derived in the course of the paper directly translate into restrictions of the underlying geometry seen by fields. 
In section \ref{sec_hyperbolicity}, we will explain that a necessary condition for the matter field equations to be predictive is that the cotangent bundle function $P$ be hyperbolic in each cotangent space. Here we will make extensive use of the theory of hyperbolic polynomials and prove two important technical lemmas. One of these will ensure that for the set of roots of reduced hyperbolic polynomials a real version of Hilbert's Nullstellensatz holds, which will be of great technical importance later. 
In section \ref{sec_masslessduals}, our focus changes from cotangent space to the geometry which a massless dispersion relation induces on tangent space. In particular, we associate vector duals with massless momenta. A central r\^ole is played here by what algebraic geometers call the dual polynomials with respect to those induced by $P$. The physical significance is that the dual polynomials emerge as the tangent space geometry seen by massless point particles, but not massive point particles as we will see later.
In section \ref{sec_bihyperbolicity}, we will see that not only the polynomials defined by $P$, but also their dual polynomials, must be hyperbolic. This is our second algebraic condition on dispersion relations. In conjunction with the third and final algebraic condition, namely that the polynomials be energy-distinguishing, this is necessary in order to have a well-defined notion of positive particle energy on which all observers agree. Thus at the end of this section, we identified all three algebraic conditions for a cotangent bundle function $P$ to serve as a physically viable massless dispersion relation. 
In section \ref{sec_massive}, we extend the theory to massive dispersion relations. Bi-hyperbolicity \new{and the energy-distinguishing property, previously recognized as essential properties} in the context of massless dispersion relations, is shown to also play a crucial r\^ole when discussing massive matter. In particular, \new{they imply an reverse triangle and an inverse Cauchy-Schwarz inequality and, above all, ensure} the existence of the Legendre duality theory for massive covectors employed in the succeeding section. 
In section \ref{sec_legendre}, the action for free massive point particles is analyzed. Compared to the dual tangent space geometry seen by massless particles, the dual geometry seen by massive particles is encoded in another, generically non-polynomial, tangent bundle function. \new{The generic emergence of two different tangent bundle geometries is an important point,} since it obstructs any attempt to devise a Finslerian extension of Lorentzian geometry which would be able to capture the massive and massless particle behaviour in one single tangent bundle structure. 
In section \ref{sec_E(p)}, we show why the decidedly covariant discussion of dispersion relations, which we adopted throughout this work, is required even if one ultimately prefers to represent the dispersion relation in form of a function $E(\vec{p})$, expressing the energy of a particle in terms of its spatial momentum. \new{This will be seen to be the case because the required temporal-spatial split of a covariant particle momentum is only defined in terms of the covariant dispersion relation}. Pushing the theory of observers and frames further, we identify the generically non-linear parallel transport induced by a bi-hyperbolic and energy-distinguishing dispersion relation and thus succeed in defining inertial laboratories.
In section \ref{sec_energycut}, we put our understanding of massive and massless momenta to work. By using virtually the entire machinery developed before, we reveal a generic \new{high-energy} effect for general dispersion relations. In particular, we determine the maximum energy a massive particle can have without radiating off, sooner or later, a massless particle in non-Lorentzian geometries. The geometric picture reveals that this is a fully covariant feature induced by the spacetime geometry. 
\revise{In section \ref{sec_pospelov}, we illustrate in detail how easy it is to check whether concrete field equations possess a bi-hyperbolic and energy-distinguishing dispersion relation. Remarkably, the popular deformations of Maxwell theory by Gambini-Pullin or Myers-Pospelov do not pass this test and are thus recognized to be non-predictive and to obstruct a well-defined notion of positive energy.} 
Finally, in section \ref{sec_conclusions}, we draw conclusions from what has been learnt, indicate limitations of the results we obtained, and point out remarkably interesting issues that one may now study based on the results of the present work.\\

Throughout the paper, the abstract theory is illustrated by showing how the constructions work out for the familiar example of metric geometry on the one hand, and for area metric geometry as a prototypical example of a non-metric geometry, where our general techniques come into full play, on the other hand. Of course, and this is one purpose of this article, the reader may instead study his own favourite candidate for a spacetime geometry or dispersion relation by the techniques developed in this paper. The results presented here apply universally.     

%\newpage
\Section{Covariant dispersion relations I: \quad Massless particles}
\label{sec_covdisprelmassless}
{\it The dispersion relation for massless matter is determined by the entirety of matter field equations one stipulates. We conclude in this section that the dispersion relation is encoded in a cotangent bundle function $P$ that induces a reduced homogeneous polynomial in every tangent space. }\\

To understand how a massless dispersion relation arises from matter field equations, we consider a (gauge fixed) action $S[\Phi,G]$ for a field multiplet $\Phi=(\Phi^N)$, where $G$ is an a priori arbitrary tensor field encoding the geometry of a smooth manifold $M$ on which the matter field dynamics are defined. Here we will notably not restrict attention to pseudo-Riemannian geometries, but rather allow a priori for any tensorial structure $G$ to play the r\^ole of the geometry. Indeed, it is one of the points of this article to find how the choice of $G$ is restricted if it is to provide a consistent classical spacetime structure. Matter field dynamics, however, are restricted from start to those giving rise to linear field equations, since only these can serve as test matter probing the geometry. More precisely, the \revise{manifestly local equations of motion then take the form
\begin{equation}\label{lin_field}
  D_{MN}(\partial) \Phi^N(x) \equiv \left[\sum_{n=1}^s Q_{MN}^{i_1 \dots i_n} \, \partial_{i_1} \dots \partial_{i_n}\right] \Phi^N(x)= 0, 
\end{equation}
%\begin{equation}\label{lin_field}
% \left[Q_{MN}^{i_1 \dots i_s}[G]\, \partial_{i_1} \dots \partial_{i_s} + \textrm{ lower order derivatives }\right] \Phi^N(x)  =0,
%\end{equation} 
where small latin indices range from $0$ to $\dim M-1$, and the coefficient matrices $Q$ at all orders depend only on the geometry $G$ but not on the value of the fields $\Phi^N$; note that this is really only true for genuinely linear field equations, and does not even hold for the linearization of fundamentally non-linear dynamics \new{\footnote{\new{For field equations obtained from a linearization around an exact solution $\Phi_0$ of genuinely non-linear dynamics, the coefficient matrix $Q^{i_1 \dots i_s}$ will be a functional not only of the geometric tensor $G$ but also of the field $\Phi_0$, so that the objects constructed from it would not only depend on the geometric tensor, as one however needs for a purely geometric interpretation. Whence we restrict attention to linear test matter.}}.}  
Remarkably, the theory of partial differential equations reveals that the entire causal structure of the local field equations are encoded in the principal polynomial $P$ associated with the equations (\ref{lin_field}). The latter is defined as the leading order term $P(q)$ of the scalar function $\omega_G \det D_{MN}(i q)$ on covectors $q$, where $\omega_G$ is some scalar density of appropriate weight, constructed from the tensor $G$ such that $P$ is a polynomial function in each cotangent fibre. We will see in the following sections that the choice of the scalar density $\omega_G$ does not affect the dispersion relation for massless matter, but that it is relevant for the dispersion relation for massive matter. 

 Concrete illustrations of this truly simple construction are given at the end of this section, where we determine the cotangent bundle function $P$ for  abelian gauge field dynamics on metric and area metric geometry as two prototypical examples, and then again in section \ref{sec_pospelov} where we are able to show the non-physicality of other popular deformations of Maxwell theory, based on the results devoleped throughout this paper.
The central observation of this section is that $P$ is recognized to induce a homogeneous polynomial $P_x$ of some degree $\textup{deg}\, P$ in each cotanget space.}

The field equations (\ref{lin_field}) may be viewed in the geometric-optical limit whenever the leading order term of the above determinant in fact coincides with the determinant of the leading order operator of the field equations,  
\begin{equation}\label{Pdef}
    P(x,q)= \omega_G\, \det(Q^{i_1 \dots i_s}(x) q_{i_1} \dots q_{i_s})\,,
\end{equation}
which indeed is guaranteed to be the case if (\ref{Pdef}) is not identically zero. 
This construction (\ref{Pdef}) is covariant in the first place because for fixed $M, N$, the leading order coefficients $Q^{i_1 \dots i_s}_{MN}$, and indeed only those, transform like a tensor, and so the determinant is a tensor density whose weight depends on the nature of the fields $\Phi^N$ and must be countered by the weight of $\omega_G$. 
 A concrete illustration of this construction is given at the end of this section, where we determine the cotangent bundle function $P$ for  abelian gauge field dynamics on metric and area metric geometry as two prototypical examples.
As we will show now, the  physical r\^ole of $P$ is that it provides the massless dispersion relation that arises in the geometric optical limit of (\ref{lin_field}), in form of the solvability condition
\begin{equation}\label{dispersion}
   P(x,q) = 0\,.
\end{equation}
This is seen by considering matter solutions in the short-wave approximation \cite{egorov1994},\cite{perlick2000},\cite{rauch},\cite{audretsch1990}, where one considers solutions of (\ref{lin_field}) taking the form of the formal series
\begin{equation}
\label{eikonalsolution}
\Phi^N(x, \lambda)= e^{i \frac{S(x)}{\lambda}} \sum^\infty _{j=0}  \phi^N_j(x)  \lambda^j
\end{equation}
and then obtains an approximate solution taking the limit $\lambda \rightarrow 0$. In the above expansion, $\phi^N_j(x)$ is a tuple of functions for each $j$, and  the scalar function $S(x)$ is known as the eikonal function. Substituting the formal series (\ref{eikonalsolution}) into the field equations (\ref{lin_field}), one finds
\begin{equation}
\label{firstnontrivialterm}
e^{i\frac{S(x)}{\lambda}} \lambda^{-s} \left[ {Q_{MN}(x)}^{i_1 \cdots i_s} \partial_{i_1} S \cdots \partial_{i_s} S\, \phi^N_0 (x) +\sum^\infty _{j=1}{v_M}_j(x)\lambda ^j \right ] = 0\,,
\end{equation}
where each of the ${v_M}_j(x)$ terms depends on some of the matrix coefficients of the differential equation (\ref{lin_field}), on  coefficients $\phi_j^N(x)$ of the expansion (\ref{eikonalsolution}) and on the eikonal function $S$ and its derivatives of \new{} lower than the highest order $s$. For $\Phi^N(x)$ to be a solution after any truncation of the series (\ref{firstnontrivialterm}), the latter has to vanish order by order in $\lambda$. Clearly, the first term $e^{i\frac{S(x)}{\lambda}} {Q_{MN}(x)}^{i_1 \cdots i_s} \partial_{i_1} S \cdots \partial_{i_s} S\, \phi^N_0 (x)$, corresponding to the power $ \lambda^{-s}$, vanishes with non-trivial $\phi_0^N$ only if the eikonal function $S$ satisfies the differential equation
\begin{equation}
\label{eikonalequation}
P(x,\partial S) \sim \textup{det}\left( {Q(x)}^{i_1 \cdots i_s} \partial_{i_1} S \cdots \partial_{i_s} S\right )=0\,,
\end{equation}
 where $P(x,\partial S)$ is recognized to be a homogeneous polynomial in $\partial S$. \new{} Equation (\ref{eikonalequation}) is known as the eikonal equation and  represents the solvability condition for the first term in (\ref{firstnontrivialterm}). Considering the first term in (\ref{firstnontrivialterm}) as an approximate solution for (\ref{lin_field}) is what is called the geometric optical limit and its \new{further relevance, beyond the r\^ole it plays for us here,} is that having this lower order approximate solution, one can generate higher order approximate solutions.
More details about the short-wave approximation, albeit only for scalar fields or otherwise only first order equations, can be found in the books by Perlick \cite{perlick2000}, Egorov and Shubin \cite{egorov1994} and the lecture notes by Rauch \cite{rauch}. In \cite{audretsch1990} equations of the type (\ref{lin_field}) are discussed.

In case $P$ is a reducible polynomial in each fibre, i.e., a product $P(x,q)=P_1(x,q)^{a_1}\cdots P_f(x,q)^{a_f}$ of irreducible \new{\footnote{\new{A non-constant real polynomial is irreducible if it cannot be written as a product of two non-constant polynomials. There is no known algorithm to decide the irreducibility of real polynomials in several real variables; a case by case analysis is required.}}} factors $P_1, \dots, P_f$ with positive integer exponents $a_1, \dots, a_f$, subtleties arise. In that case \cite{egorov1994},\cite{perlick2000} one has to take as the \new{cotangent bundle function $P$} the reduced polynomial
 \begin{equation}
 \label{reducedpolynomial}
P(x,q)=P_1(x,q)\cdots P_f(x,q),
\end{equation} 
in other words, one must remove repeated factors in the original polynomial. We will have more to say about the relation between properties of a reduced polynomial and those of its individual factors in section \ref{sec_hyperbolicity}. 
\begin{figure}[h]
\includegraphics[width=16cm,angle=0]{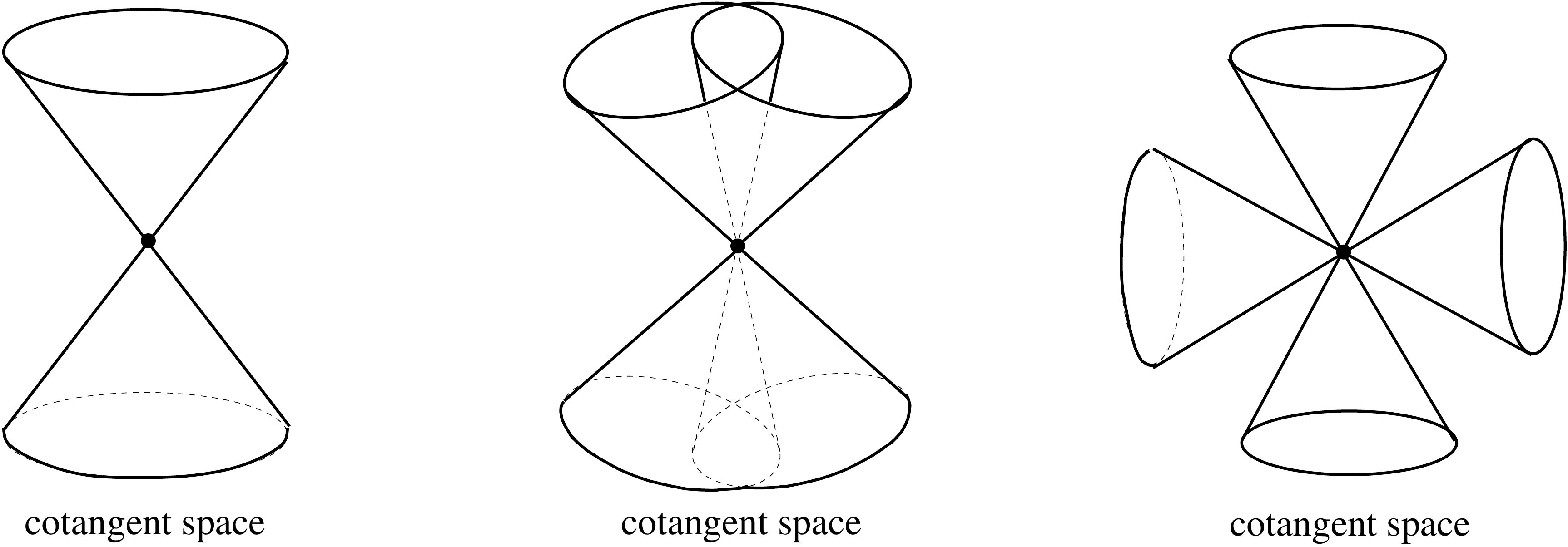}
\caption{\label{fig_homogeneous} Homogeneity of $P$ in its cotangent fibre gives rise to a cone of $P$-null covectors in each cotangent space. Three prototypical examples of such cones are shown, the first one being the familiar Lorentzian metric cone. \new{Only for the second example do we have $N_x \neq N_x^\textrm{smooth}$, with the difference set being constituted by the covectors lying in the intersection of the cones.}}
\end{figure}

Clearly, this cancelation does not alter the set \new{$N_x$} of massless momenta at a spacetime point $x$, which due to the homogeneity of $P_x$ constitutes the algebraic cones 
\begin{equation}
  N_x = \{k \in T_x^*M \,|\, P(x,k)=0 \}\,,
\end{equation}
which \new{is to say} that every positive real multiple of a massless momentum is again a massless momentum, \new{cf.} figure \ref{fig_homogeneous}. For technical precision, we will occasionally focus on the smooth subcone   
\begin{equation}
  N^\textrm{smooth}_x = \{ k \in  N_x \,|\, DP(x,k)\neq 0\}\,, 
\end{equation} 
where $DP$ denotes the derivative of $P$ with respect to the cotangent fibre. So the cotangent bundle function $P$ determines the (smooth) massless momentum cone.
 The converse question, namely under which conditions the massless momentum cone \new{$N_x$} at a point $x$ determines the polynomial $P_x$ up to a constant factor, is subtle, but of central importance. The vanishing sets associated with polynomials are the subject of study of algebraic geometry, and we will indeed have opportunity to employ some elaborate theorems of real algebraic geometry. In the remainder of this first section, we clarify the relation between vanishing sets of real polynomials and the principal ideals that these polynomials generate\new{, since this will be relevant later}. Recall that an ideal $I\subset R$ in a ring $R$ (where $R$ is here concretely the ring of real polynomials in $\dim M$ real variables) is a subset that is closed under addition and under multiplication with an arbitrary ring element.  Now on the one hand, we may consider the situation where we are given an ideal $I$ and define the vanishing set $\mathcal{V}(I)$ as the set of cotangent vectors that are common zeros to all polynomials in $I$. On the other hand, we may be given a subset $S$ of cotangent space and consider the set $\mathcal{I}(S)$ of all polynomials in $R$ that vanish on all members of that set $S$. Now it can be shown that $\mathcal{I}(S)$ is an ideal in the ring of polynomials on cotangent space, and that one always has the inclusion
\begin{equation}
  \mathcal{I}(\mathcal{V}(I)) \supseteq I\,.
\end{equation}
The question under which conditions equality holds is studied in the Nullstellens\"atze of algebraic geometry. While this is a relatively straightforward question for polynomials over algebraically closed fields \cite{hassett2007}, such as the complex numbers, \new{for the real numbers underlying our study here, one needs to employ a string of theorems that were originally developed in order to solve Hilbert's seventeenth problem.}
Indeed, for a reduced homogeneous polynomial $P_x$\new{,} one obtains the equality 
\begin{equation}\label{realnullstellen}
  \mathcal{I}(N_x) = \langle P_x \rangle\,,
\end{equation}     
if $N^\textrm{smooth}_x \neq \emptyset$ (which condition we will be able to drop for the hyperbolic polynomials \new{to which we will narrow our attention from the next section onward}, see the remarks following the First Lemma proven there). Here $\langle P_x \rangle$ denotes the ideal containing all polynomials that have $P_x$ as a factor. Drawing on the said results from real algebraic geometry, this is seen as follows. Let $P_{x\,i}$ be the $i$th irreducible factor of $P_x$
Then there exists a $q \in N^\textrm{smooth}(P_{x\,i})$ so that corollary 2.9 of \cite{dubois1970} shows that $P_{x\,i}$ generates a real ideal, i.e., $\mathcal{I}(N(P_{x\,i})) = \langle P_{x\,i} \rangle$. According to corollary 2.8 of \cite{dubois1970}, the reduced polynomial $P_x$ thus also generates a real ideal since it does not contain repeated factors. Finally theorem 4.5.1 of \cite{bochnak1998} yields the claim. The equality (\ref{realnullstellen}) will play a significant technical r\^ole in ensuring that we can determine the vector duals of massless momenta using elimination theory, in section \ref{sec_masslessduals}.\\

\subsection{Illustration: Maxwell theory on metric and area metric backgrounds}
We now illustrate, by way of two concrete examples, how the cotangent bundle function $P$ and thus the massless dispersion relation are extracted from a field theory on a given geometry.

Our first example, Maxwell theory on a metric background, is of course a classic problem, and developed in full detail for instance in \cite{perlick2000}. Here we present only those results which are illuminating with regards to the present work. So let $(M,g)$ be a metric manifold of arbitrary signature, and  consider a one-form field $A$ whose dynamics is governed by its coupling to the metric tensor $g$ according to the Maxwell action
\begin{equation}
\label{metricmaxwellaction}
 S[A,g]=-\frac{1}{4}\int \,d^4x\,\sqrt{\mid\textup{det}(g)\mid} \, g^{am} g^{bn} F_{mn} F_{ab},
\end{equation}
where $F=dA$ is the field strength. Moreover, an orientable metric manifold carries a canonical volume form ${\omega_g}_{abcd}=\mid \textup{det}(g)\mid ^{ 1/2} \epsilon_{abcd}$, so that after variation of the action (\ref{metricmaxwellaction}) with respect to the one-form field $A$, and after introducing coordinates $x^a=(t,x^\alpha)$, one can rewrite the obtained second order field equations for $A$ as a system of first order field equations (plus two constraint equations) for the electric field  $ E_{\alpha}=F(\partial_t,\partial_\alpha)$ and the magnetic field $B^{\alpha}=\omega^{-1}_g(dt,dx^\alpha,F)$ as
\begin{equation}
\label{firstorderpde}
\left( {{A^b}^M}_N \,\partial _b +{B^M}_N\right) u^N=0,
\end{equation}
where $u^N=(E_\alpha,B^\alpha)$ and the matrices ${{A^b}^M}_N$ depend on the metric tensor $g$ and the volume form $\omega_g$, see \cite{perlick2000} for the exact dependence. Strictly speaking, choosing the field strengths as dynamical variables does away with the gauge symmetry not precisely by gauge-fixing the action, as was assumed in the general part of this section, but achieves the same at the level of the equations of motion and was chosen because it allows for a more concise discussion here. In any case, the system (\ref{firstorderpde}) is a particular case of (\ref{lin_field}), so that the massless dispersion relation is given by some scalar density $\tilde P
_x (q)=\textup{det}({A^b} q_b)=0$. Inserting the explicit expression for ${A^b}^M{}_N$, one finds $\tilde P_x (q)=q_0^2 (g_x ^{-1}(q,q))^2$. To obtain the massless dispersion relation, we must cancel repeated factors in $\tilde P_x(q)$, as explained above. Moreover, one can show \cite{rauch} that $q_0=0$ is inconsistent with the constraint equations, so that  we finally obtain the massless dispersion relation on a metric background as derived from Maxwell field equations as $P_{x}(q)=g_x ^{-1}(q,q)=0$, the familiar result.

As a second example, we discuss area metric geometry. We provide only the most basic definitions and results needed for the purpose of this example; for a more detailed introduction, see e.g. \cite{schuller2005},\cite{schuller2006},\cite{schuller2010}. An area metric manifold $(M, G)$ is a smooth differentiable manifold $M$ equipped with a
smooth covariant fourth rank tensor field $G$ with the algebraic symmetries $G_{abcd} = G_{cdab}$ and $G_{abcd} = G_{bacd}$. Moreover, an area metric is required to be invertible in the sense that there is a smooth tensor $G^{abcd}$
so that $G^{abpq} G_{pqcd} = 2(\delta ^a _c \delta ^b _d  - \delta^a_d \delta ^b _c )$. In particular, an area metric $G$ can be viewed as a metric $\textrm{Petrov} \left( G \right)$ in the space of two forms and it induces the canonical volume form $(\omega_G)_{abcd}= (\textup{det}\left(  \textrm{Petrov} \left( G \right) \right))^{1/6} \epsilon_{abcd}$ for $\dim M=4$.
Maxwell theory on an area metric background was fully studied in \cite{punzi2009},\cite{schuller2010} and here we only summarize the main results. We consider a one-form field $A$ coupled to an area metric background according to the action
\begin{equation}
\label{areametricaction}
 S[A,G]=-\frac{1}{8}\int \, d^4 x \mid \textup{det}\left(   \textrm{Petrov} \left( G \right) \right) \mid^{1/6} F_{ab} F_{cd}\, G^{abcd}\,,
\end{equation}
where $F=dA$ is the field strength. After variation of the action (\ref{areametricaction}) with respect to the one-form field $A$ and after introducing coordinates $x^a=(t,x^\alpha)$ one can rewrite, as in the metric case, the obtained second order field equations for $A$ as the system (\ref{firstorderpde}) of first order field equations (plus two constraint equations). For this case $u^N=(E_\alpha,B^\alpha)$ with  electric field $ E_{\alpha}=F(\partial_t,\partial_\alpha)$ and magnetic field $B^{\alpha}=\omega^{-1}_G(dt,dx_\alpha,F)$, and the matrices ${{A^b}^M}_N$ depend now on the area metric tensor $G$ and the volume form $\omega_G$. For the explicit dependence see \cite{schuller2010}.  Then the massless dispersion relation must again be given by $\tilde P_x(q)=\textup{det}({A^b}\, q_b)=0$. After an explicit calculation one finds $\tilde P_x(q)=q_0^2 P_{x}(q)$  with
\begin{equation}
\label{fresnelpolynomial}
P_{x}(q)=-\frac{1}{24}(\omega_{G_x})_{mnpq} (\omega_{G_x})_{rstu}  G_x^{mnr(a} G_x^{b\mid ps \mid c} G_x^{d)qtu}q_a q_b q_c q_d\,.
\end{equation}
 One can also show, as in the metric case, that $q_0=0$ is inconsistent with the constraint equations, such that  we finally find that the massless dispersion relation on an area  metric background as derived from Maxwell field equations is given by $P_{x}(q)=0$. This result has been obtained first by Hehl, Rubilar and Obukhov \cite{Obukhov2002},\cite{Hehl2002} in the context of premetric electrodynamics.

%\newpage
\Section{Hyperbolicity}\label{sec_hyperbolicity}
{\it Employing our knowledge on how a cotangent bundle function $P$ encoding the massless dispersion relation arises from field equations, we now identify the hyperbolicity of $P$ as a crucial condition for the field equations to be predictive in the first place. For the cone of massless momenta, the real Hilbert Nullstellensatz is thus shown to hold without further conditions.}\\

The second property required for a cotangent bundle function $P$ to provide a viable massless dispersion relation, besides being a reduced homogeneous polynomial in the fibre coordinate, also originates in the underlying matter field equations (\ref{lin_field}) and goes right to the heart of what classical physics is all about. Namely that the theory be predictive. In other words, initial data on a suitable initial data surface are required to evolve in a unique manner. Remarkably, the question of what constitutes suitable initial data surfaces on the one hand, and the question of whether the field equations evolve the initial data in a unique way on the other hand, are both decided by an algebraic property of the cotangent bundle function $P$. More precisely, an inescapable condition for the initial value problem to be well-posed in a region of spacetime is that $P$ defines a hyperbolic polynomial $P_x$ at every point $x$ of this region \cite{hoermander1977},\cite{ivrii1974}.
\begin{figure}[h]
\includegraphics[width=10cm,angle=0]{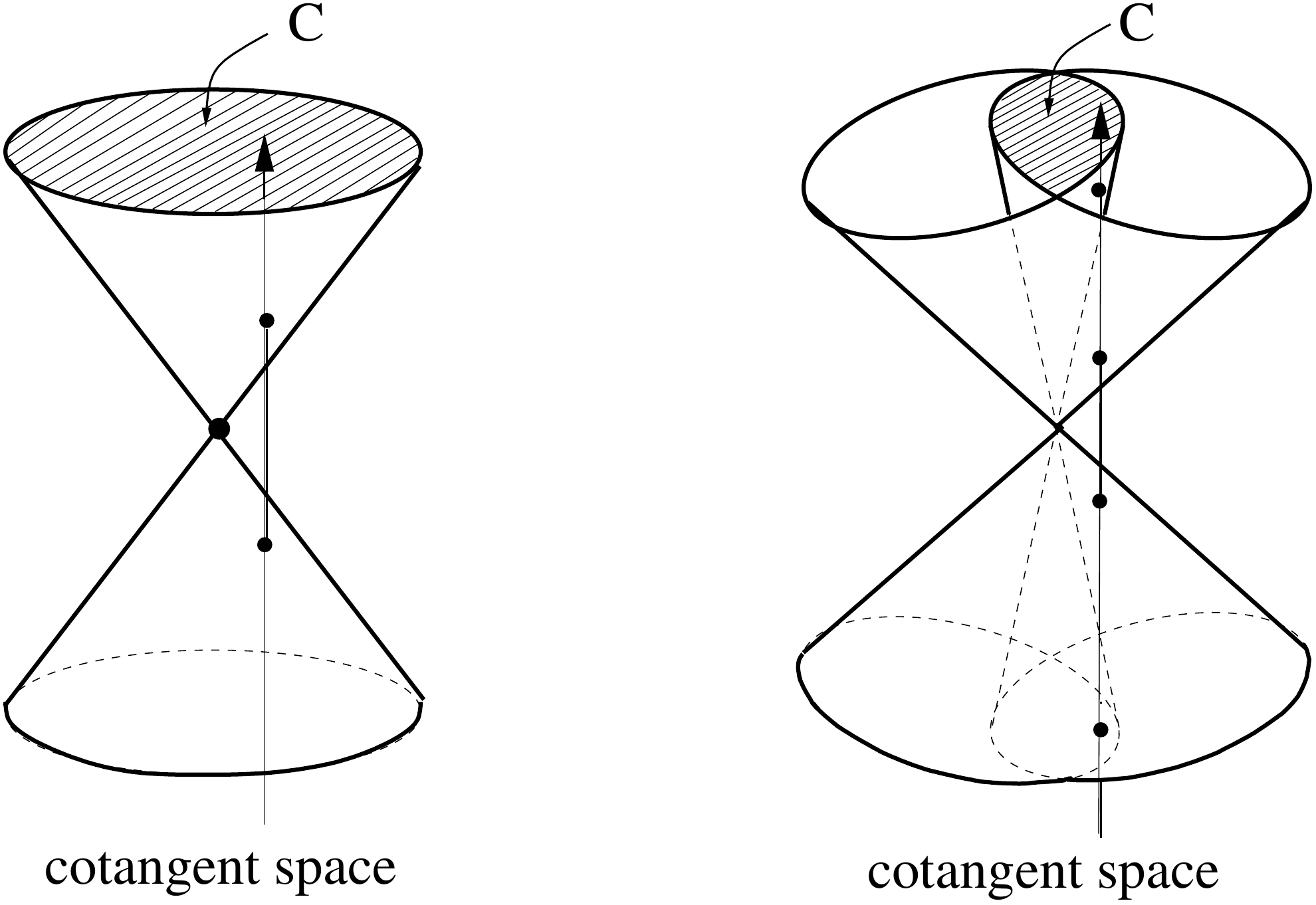}
 \caption{\label{fig_hyperbolic} Hyperbolicity cones for two prototypical polynomials. On the left the familiar second degree Lorentzian cone; on the right a fourth degree cone defined, for simplicity, by a product of two Lorentzian metrics.}
\end{figure}
A homogeneous polynomial $P_x$ is called hyperbolic with respect to some covector $h$ if for every covector $q$ with $P_x(q)\neq 0$ any $\lambda$ solving 
\begin{equation}\label{defcone_local}
  P_x(q + \lambda h) = 0
\end{equation}
is a real number. This definition of hyperbolicity is easy to understand in geometric terms. It simply means that there is at least one covector $h$ such that every affine line in cotangent space in the direction of $h$ intersects the cone defined by (\ref{dispersion}) in precisely $\deg P$ points, see figure \ref{fig_hyperbolic}, counting algebraic rather than geometric multiplicities. Any such covector $h$ identifying $P_x$ as hyperbolic is itself called a hyperbolic covector at the point $x$. The various connected sets of hyperbolic covectors in the same cotangent space, as for instance the upper (shaded) cones in figure \ref{fig_hyperbolic}, are called the hyperbolicity cones of $P$ at $x$.  It is often useful to take a more global point of view and consider a smooth distribution $C$ of hyperbolicity cones $C_x$ over all spacetime points $x$, which one simply may think of as the cone of all smooth covector fields $h$ for which $h_x\in C_x$. More precisely, let $h$ be a covector field hyperbolic with respect to $P$, that is $h$ defines a hyperbolic covector at every spacetime point. Then the hyperbolicity cone $C(P,h)$ containing $h$ is constituted by all covector fields $q$ with the property that all functions $\lambda$ on $M$ satisfying
\begin{equation}
\label{defcone}
  P(x,q(x)-\lambda(x) h(x)) = 0
\end{equation}
are positive everywhere on $M$. The cone $C(P,h)$ induces a cone $C_x(P,h)$ in each cotangent space $T_x^*M$, consisting of the values $q(x)$ of all $q\in C(P,h)$ evaluated at $x$, which is called the hyperbolicity cone of $P$ with respect to $h$ at $x$. Clearly, $C_x(P,h)$ only depends on the value of $h$ at $x$, and thus one may think of $C(P,h)$ simply as the said distribution of the $C_x(P,h)$ over all $x \in M$. 
The somewhat implicit definition of hyperbolicity cones, both the local (\ref{defcone_local}) and the global one (\ref{defcone}), can be cast into the explicit form of $\deg P$ polynomial inequalities, as was proved in \cite{gueler1997} from the Routh-Hurwitz theorem, if at least one hyperbolic covector within the cone one desires to describe is known. More precisely, let $P$ be hyperbolic with respect to $h$ so that $P(x,h(x))>0$ without loss of generality. Then the hyperbolicity cone is described by the $\deg P$ inequalities
\begin{equation}
  \det H_i(v,h) > 0\qquad \textrm{ for all } i=1,\dots,\deg P\,, 
\end{equation}
where the matrices $H_1, H_2, \dots, H_{\deg P}$ are constructed as 
\begin{equation}
  H_i(v,h) = \left[\begin{array}{ccccc}
    h_1 & h_3 & h_5 & \dots & h_{2i-1} \\
    h_0 & h_2 & h_4 & \dots & h_{2i-2} \\
    0   & h_1 & h_3 & \dots & h_{2i-3} \\
    0   & h_0 & h_2 & \dots & h_{2i-4} \\
    \vdots  &  \vdots  &  \vdots  &  \vdots  &  \vdots\\
    0 & 0 & 0 & \dots & h_i 
\end{array}\right]_{i \times i} \qquad \textrm{where } h_j \textrm{ is set to } 0 \textrm{ for } j>i
\end{equation}
from the coefficients of the expansion 
\begin{equation}
  P(x,v + \lambda h) = h_0(x,v,h)\, \lambda^{\deg P} + h_1(x,v,h)\, \lambda^{\deg P - 1} + \dots + h_{\deg P}(x,v,h)\,.
\end{equation}
Similar to algebraic geometry dealing with algebraic sets defined by polynomial equations, there is a rather elaborate theory of semi-algebraic sets \cite{benedetti1990},\cite{bochnak1998} defined by polynomial inequalities, of which the hyperbolicity cones are, according to the above theorem, a particular instance. Employing this theory will be of advantage in the proof of the first lemma, below.
The remarkable properties of hyperbolicity cones, which underlie all further constructions, have been elucidated by G\aa rding \cite{garding1959} \new{a long time ago}. Recalling that a subset $C$ of a real vector space $V$ is called a convex cone if \new{besides any real multiple sof an element of $C$ also the sum of any two elements of $C$ lies again in $C$}, G\aa rding proved the following results. First, any covector field $b$ belonging to a given hyperbolicity cone $C(P,a)$ equally represents the hyperbolicity cone, $C(P,b) = C(P,a)$. Second, $C(P,a)$ is an open and convex cone. Third, $P$ is strictly non-zero on $C(P,a)$, but vanishes on the boundary $\partial C(P,a)$. \new{Finally}, the suitable initial data surfaces, alluded to at the beginning of this section, are recognized as those whose normal covector fields are hyperbolic. 

The calculation of hyperbolicity cones is significantly simplified if the cotangent bundle function $P$ is factorizable into factors of multiplicity one, as in eq. (\ref{reducedpolynomial}). Then P is hyperbolic with respect to $h$ if and only if each of its individual factors is hyperbolic with respect to $h$. For such reducible $P_x$, the determination of the hyperbolicity cone with respect to some hyperbolic covector $h$ is reduced to the determination of the hyperbolicity cones of the individual factors, since 
\begin{equation}\label{capcones}
  C_x(P_x, h_x)  = C_x(P_{1\, x},h_x) \cap \dots \cap C_x(P_{f\, x},h_x)\,.
\end{equation}
Thus it is not a coincidence that the hyperbolicity cone indicated in figure \ref{fig_hyperbolic} is the intersection of the hyperbolicity cones of the two Lorentzian cones whose union constitutes the vanishing set of the underlying dispersion relation. We conclude this section with the proof of two key properties of hyperbolic polynomials, which we will use repeatedly throughout this paper. \\

\noindent{\bf First Lemma.} For a reduced homogeneous hyperbolic polynomial $P_x$, the set $N_x^\textrm{smooth}$ is a dense subset of the cone $N_x$ of massless momenta.\\

{\it Proof.} 
Since our variety $N$ is generated from a single polynomial $P$, i.e., $\langle P \rangle = \mathcal{I}(N)$, it follows from Definition 3.3.4 of \cite{bochnak1998} that the set of singular points is $\textrm{Sing}(N) = N\setminus N^\textrm{smooth}$. But then $\dim \textrm{Sing}(N) < \dim N = \dim M - 1$, where the inequality is Proposition 3.3.14 of \cite{bochnak1998} and the equality follows from the hyperbolicity of $P$ \cite{beig2006}. Thus we know that the singular set is at most of dimension $\dim M - 2$. Further, we know from the first remark in 3.4.7 of \cite{benedetti1990} that $\textrm{Sing}(N)$, being a real algebraic set, can be expressed as a finite union of analytic semialgebraic manifolds $S_i$ and that every such manifold has a finite number of connected components. From the propositions 2.8.5 and 2.8.14 of \cite{bochnak1998} we thus obtain that $\dim \textrm{Sing}(N)=\max(\dim(S_i))=\max(d(S_i))$, where $d(A_i)$ is the topological dimension of the semialgebraic submanifold $S_i\subset T_x^*M$. Since $\dim \textrm{Sing}(N) \le \dim M - 2$ we conclude that $\textrm{Sing}(N)$ consists of only finitely many submanifolds of $\mathbb{R}^n$ of topological dimension less or equal to $\dim M -2$. Thus its complement $N^{\textrm{smooth}} = N\setminus \textrm{Sing}(N)$ is dense in $N$.\\
 
This property of hyperbolic polynomials is also mentioned in \cite{beig2006}. As an important corollary we obtain that the real Nullstellensatz (\ref{realnullstellen}) holds for any reduced hyperbolic polynomial without further conditions\new{; for} if $P$ is hyperbolic, certainly $N_x$ is a non-empty set of codimension one, so that the dense subset $N^\textrm{smooth}_x$ must be non-empty.\\  

\noindent{\bf Second Lemma.} If $P_x$ is a reduced homogeneous hyperbolic polynomial with hyperbolicity cone $C_x$ at some point $x\in M$ then for all covectors $s \in T_x^*M \backslash\textrm{closure}(C_x)$ there exists a massless covector $r$ on the boundary $\partial C_x$ of the hyperbolicity cone such that $s(D P_x(r))<0$.\\

{\it Proof.} It is clear that if $y \in C_x$ and $s \not\in \textrm{closure}(C_x)$, the line $y+\lambda s$ intersects the boundary $\partial C_x$ at some $r_0 = y + \lambda_0 s$ for some positive $\lambda_0$. Thus $P_x(r_0)=0$ and, since $P_x(C_x)>0$, we have $P_x(r_0-\epsilon s)>0$ for sufficiently small positive $\epsilon$. Now we must distinguish two cases: First assume that $P_x(r_0+\epsilon s) <0$, from which it follows that   
 $\frac{d}{d\epsilon}P_x(r_0+\epsilon s)|_{\epsilon=0}=s(D P_x(r_0))<0$, which proves the lemma with $r:=r_0$; Second, assume that $P_x(r_0+\epsilon s) > 0$ which is equivalent to $\frac{d}{d\epsilon}P_x(r_0+\epsilon s)|_{\epsilon=0}=s(D P_x(r_0))=0$ which in turn holds if and only if $D P_x(r_0)=0$ (to see the latter equivalence assume that, to the contrary, $s(D P_x(r_0))=0$ and $D P_x(r_0)\neq0$; this implies that $s$ must be tangential to $\partial C_x$ at $r_0$, but since $y$ lies in $C_x$ and $C_x$ is a convex cone $y+\lambda s$ could then not intersect $\partial C_x$ at $r_0$, which we however assumed). So to prove the lemma in this second case, we need to construct another $r_0' \in \partial C_x$ that satisfies the condition $s(DP_x(r_0'))<0$. Now since the First Lemma guarantees that the set $N_x^{\textrm{smooth}}$, on which $DP_x$ is non-zero, lies dense in $N_x$, we can find in every open neighborhood $U$ around $r_0$ a vector $r_0^{'}\in\partial C_x$ such that $D P_x(r_0^{'})\neq 0$. We define $z:=r_0^{'}-r_0$ and $y':=y+z$. Since $C_x$ is an open cone, $y^{'}$ lies in $C_x$ if we choose the neighborhood small enough, and the line $y'+\lambda s$ intersects $\partial C_x$ at $r_0^{'}$. Finally since $r_0^{'}\in\partial C_x$ we know that $P_x(r_0^{'})=0$ and $P_x(r_0^{'}-\epsilon s)>0$. We conclude that $s(D P_x(r_0^{'}))<0$. This proves the second lemma with $r:=r_0^{'}$. \\

%\newpage
\Section{Vector duals of massless momenta: Gauss map}\label{sec_masslessduals}
{\it So far, our considerations have focused on the geometry that is impressed by the massless dispersion relation on each cotangent space. In this section, we now associate vector duals in tangent space with the massless momenta defined by the dispersion relation. A central r\^ole is played by \new{a} dual polynomial $P^\#_x$ on tangent space that is associated with the polynomial $P_x$ on the corresponding cotangent space. \new{Physically, the} dual polynomial emerges as the tangent space geometry seen by massless point particles.}\\

In order to associate velocity vectors with massless particle momenta \new{in physically meaningful fashion}, we employ the dynamics of free massless point particles. Their dynamics, in turn, are uniquely determined by the dispersion relation, because the Helmholtz action
\begin{equation}\label{Helmmassless}
  I_0[x,q,\lambda] = \int d\tau \left[q_a \dot x^a + \lambda P(x,q)\right]
\end{equation} 
describes particles that are free due to the form of the first term and massless because of the Lagrange multiplier term. In the following, we wish to eliminate the momentum $q$ and \new{the Lagrange multiplier $\lambda$ to} obtain an equivalent action in terms of the particle trajectory $x$ only. Variation of the Helmholtz action with respect to \new{} $\lambda$ of course enforces the null condition for the particle momentum. Now variation with respect to $q$ yields $ \dot x = \lambda \, DP_x(q)$ for all $q\in N^\textrm{smooth}$,
 which implies the weaker equation
\begin{equation}\label{noninvertible}
  [DP_x(q)] = [\frac{\dot x}{\lambda}]\,,
\end{equation} 
where $[X]$ denotes the projective equivalence class of all vectors collinear with the vector $X$. 

In order to \new{solve (\ref{noninvertible}) for $q$, we need} the inverse of the projective map $[DP]$.  
\begin{figure}
\includegraphics[width=13cm,angle=0]{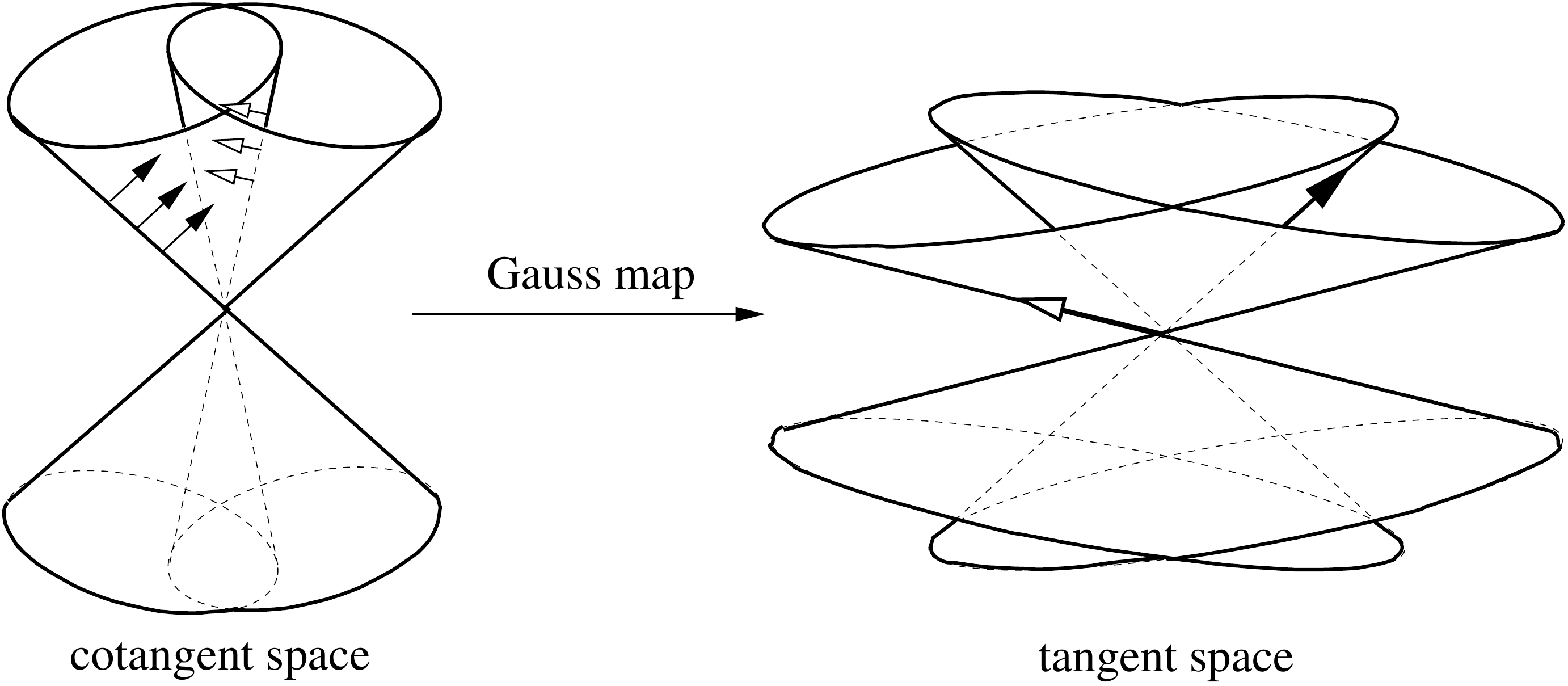}
\caption{\label{fig_dualconesymm}Gauss map sending the zero locus of a polynomial to the zero locus of a dual polynomial}
\end{figure}
We will now derive that this inverse is given by the gradient of \new{a} so-called dual polynomial.
Indeed, the image $N_x^\#$ of the massless covector cone $N_x$ under the gradient map $DP$ is again described by a homogeneous polynomial $P_x^\#$, albeit of generically different degree than $P$.
More precisely, for an irreducible cotangent bundle function P, we look for a likewise irreducible tangent bundle function $P^\#$ that is uniquely determined up to a real constant factor at each point $x$ of the manifold by the equation
\begin{equation}
\label{dualpolynomial}
  P_x^\#(DP_x(N^\textrm{smooth}_x))=0\,.
\end{equation} 
The polynomials $P_x$ and $P_x^\#$ given by $P$ and $P^\#$ at any given point $x$ of the base manifold are \new{then} called dual to each other, and it is convenient to also call the corresponding cotangent bundle function $P$ and tangent bundle function $P^\#$ dual to each other.
For a cotangent bundle function $P$ that is reducible into irreducible factors
\begin{equation}
P(x,k)=P_1(x,k)\cdots P_f(x,k)\,,
\end{equation}
we define the dual tangent bundle function as the product
\begin{equation}
P^\#(x,v) = P_1^\#(x,v) \cdots P_f^\#(x,v)\,,
\end{equation}
where the $P_i^\#$ are the irreducible duals of the irreducible $P_i$ determined by equation (\ref{dualpolynomial}). Thus $P^\#$ is uniquely determined up to a real factor function on $M$ and satisfies again equation (\ref{dualpolynomial}), as one easily sees from application of the product rule.

The proof for the existence of a dual $P_x^\#$, and indeed its algorithmic computability for any reduced hyperbolic polynomials $P$, is provided by a branch of algebraic geometry known as elimination theory. This is where the real Nullstellensatz, discussed at the end of section \ref{sec_covdisprelmassless} and shown to hold for any reduced hyperbolic polynomial at the end of the previous section, becomes essential. For if the real Nullstellensatz holds, Proposition 11.10 of \cite{hassett2007} asserts that the polynomial conditions which a vector $X$ must satisfy, in order for there to be a solution to the problem of having the polynomials  
\begin{equation}\label{elimpoly}
  P_x(k),\quad X^{i_1}-DP_x(k)^{i_1},\quad \dots, \quad X^{i_d} - DP_x(k)^{i_d}
\end{equation}
all vanish for some $k$, are obtained by first calculating an elimination ideal. How this is done in practice, by using Buchberger's algorithm and Gr\"obner bases, is explained most lucidly in \cite{hassett2007}. The \new{so} calculated elimination ideal\new{, however, may turn out to} be generated by several real homogeneous polynomials. However, making use of the fact that we are dealing with real polynomials, it is easy to construct the dual $P^\#$ as a sum of appropriate even powers of the generating polynomials, which obviously vanishes where and only where all generators vanish.   
It should be said that while \new{for most polynomials of interest,} a direct calculation of dual polynomials using elimination theory exhausts the capability of current computer algebra systems \new{}, in many cases one is nevertheless able to guess the dual polynomial by physical reasoning (as we will illustrate for the cases of metric and  area metric geometry \new{at the end of this section}). Once such an educated guess has been obtained, one may directly use the defining equation (\ref{dualpolynomial}) to verify that one has found the dual polynomial. In any case, since its existence is guranteed, we will simply assume in the following that a dual $P^\#$ has been found by some method.

Equipped with the notion of the dual polynomial, we may now return to the projective gradient map
\begin{equation}
  [DP_x]: [N^\textrm{smooth}_x] \to [N_x^\#]\,,\qquad [q] \mapsto [DP_x(q)] 
\end{equation}
first encountered in (\ref{noninvertible}), where the brackets denote projective equivalence classes, identifying parallel vectors (respectively covectors), but not antiparallel ones, and $N_x^\#$ is the image of $N^\textrm{smooth}_x$ under $DP_x$. The projective map $[DP_x]$ is well-defined due to the homogeneity of $P_x$, and will be refered to as the Gauss map.
The problem of inverting the Gauss map is now solved by definition of the dual Gauss map $[DP_x^\#]$ in terms of the dual polynomial $P_x^\#$,
\begin{equation}
 [DP_x^\#]: [N_x^{\#\textrm{ smooth}}] \to [N_x]\,,\qquad [X] \mapsto [DP_x^\#(X)]\,,
\end{equation}
since we then have for null covectors $k\in N_x^\textrm{smooth}$ that 
\begin{equation}\label{DDPneq0}
 [DP_x^\#]([DP_x]([k])) = [k]  \qquad \textrm{ if }\, \det(DDP_x)(k)\neq 0\,,
\end{equation}
so that the dual Gauss map $[DP^\#]$ acts as the inverse of the Gauss map on the images of all covectors $k$ satisfying the above \new{determinantal} non-degeneracy condition.
That relation (\ref{DDPneq0}) holds is most easily seen from rewriting the duality condition (\ref{dualpolynomial}) in the form
\begin{equation}
  P^\#(x,DP(x,k)) = Q(k) P(k) \qquad \textrm{ for all covectors } k\,,
\end{equation}
since this form does not require an explicit restriction to null covectors. Thus differentiation with respect to $k$ yields, by application of the chain rule and then of Euler's theorem \new{\footnote{\new{Euler's theorem asserts the simple fact that for any function $f$ that is homogeneous of degree $\deg f$, the relation $Df(v) v=(\deg f) f(v)$ holds for any $v$ in the domain of $f$.}}} on the right hand side, for any null covector $k$ satisfying the non-degeneracy condition in (\ref{DDPneq0}) that
\begin{equation}
  \new{DP^\# (x,DP(x,k)) = \frac{Q(x,k)}{\deg P - 1} k}  \,,
\end{equation}
which in projective language takes the form (\ref{DDPneq0}).
In particular, we may thus solve the projective equation (\ref{noninvertible}) for
\begin{equation}\label{projnullmap}
  [q] = [DP_x^\#]([\dot x/\lambda])\,.
\end{equation} 
Obviously, the homogeneity of $DP_x^\#$ in conjunction with the projection brackets \new{allows} to disregard the function $\lambda$ altogether. However, another undetermined function $\mu$ appears when translating this result back to non-projective language,
\begin{equation}\label{nullmap}
  q = \mu\,  DP_x^\#(\dot x)\,.
\end{equation}   
Now we may \new{replace} the momentum in (\ref{Helmmassless}) by this expression and use \new{again Euler's theorem applied to the } homogeneous polynomial $P_x^\#$ to finally obtain \new{the massless point particle action}
\begin{equation}\label{nullaction}
  I_0[x,\mu] = \int d\tau \mu\, P^\#(x,\dot x)\,.
\end{equation}
Relations (\ref{noninvertible}) and (\ref{projnullmap}) reveal the physical meaning of the \new{Gauss map} $[DP_x]$ and \new{its inverse} $[DP_x^\#]$: up to some irrelevant conformal factor, they associate null particle momenta in $N_x^\textrm{smooth}$ with the associated null particle velocities in $N_x^{\#\textrm{ smooth}}$. The \new{automatic} appearance of a final Lagrange multiplier $\mu$ in (\ref{nullaction}) also hardly comes as a surprise, since it is needed to enforce the null constraint $P_x^\#(\dot x)=0$.
This reveals the direct physical relevance of the dual tangent bundle function $P^\#$ as \new{} the tangent space geometry seen by massless particles. 

\subsection{Illustration: Dual polynomials for metric and area metric geometry}
As an illustration of the above abstract theory, we provide the explicit form of the dual tangent bundle functions $P^\#$ associated with the cotangent bundle functions induced by abelian gauge theory on first metric and then area metric geometry. 

For a metric manifold $(M,g)$, we saw that the cotangent bundle function was given by $P_g(x,q)=g_x^{-1}(q,q)$. It is easy to guess its dual, namely $P^\#_g(x,v)=g(v,v)$. Indeed, $P^\#_{g\,x}(D P_{g\,x}(q))=4 g_x(g_x^{-1}(q,\cdot),g_x^{-1}(q,\cdot))=4 g_x^{-1}(q,q)=4 P_{g\,x}(q)$, so that equation (\ref{dualpolynomial}) is satisfied. Thus we conclude that  $P^\#_{g\,x}(x,v)=g_x(v,v)$ is the dual polynomial of $P_{g\,x}(q)$.

The case of an area metric manifold $(M,G)$, where the cotangent bundle function $P_G$ induced by abelian gauge theory is given by (\ref{fresnelpolynomial}), is already cosiderably more complicated. At first sight it could seem that there is no way to avoid  the use of elimination theory. However, already in four dimensions, elimination theory is prohibitively difficult for current computer algebra programs, even if full use is made of our knowledge of normal forms for area metrics \cite{schuller2010}. So while in principle Bucherberger's algorithm applies, practically one is better off obtaining an educated guess for what the dual polynomial might be, and then verifying that guess employing equation (\ref{dualpolynomial}). Thanks to the invertibility properties of area metrics, an educated guess for the dual of $P_{G}$ can be derived directly from Maxwell theory \cite{punzi2009}. In the language of \cite{punzi2009}, a wave covector field $q$ for an abelian gauge field field strength $F$ is a section of $T^*M$ satisfying
\begin{equation}
\label{areametric_geometricoptical}
 q\wedge F=0, \quad \quad q\wedge H=0\,,
\end{equation}
where the constitutive relation between the field strength $F$ and the induction $H$ is given as $H_{ab}=-1/4 |\text{det} G|^{1/6} {\omega_G}_{abmn} G^{mnpq} F_{pq}$. Solving (\ref{areametric_geometricoptical})  leads to the Fresnel polynomial (\ref{fresnelpolynomial}). Dually, and this is the key idea, ray vector fields $v$ on the discontinuity surface determined by a wave covector field $q$ are defined as 
\begin{equation}
\label{rayvectors}
 F(v,\cdot)=0, \quad \quad  H(v,\cdot)=0\,,
\end{equation}
where F and H are solutions of (\ref{areametric_geometricoptical}). Looking for solutions of the system (\ref{rayvectors}), one finds that the condition for their existence is that the ray vector fields $v$ must satisfy the polynomial equation
\begin{equation}
\label{areametricdualpolynomial}
P_{G\,x}^{\#}(v)=-\frac{1}{24}(\omega^{-1}_{G_{x}})^{mnpq} (\omega_{G_x}^{-1})^{rstu}  {G_x}_{mnr(a} {G_x}_{b\mid ps \mid c} {G_x}_{d)qtu}v^a v^b v^c v^d=0\,,
\end{equation}
which for physical reasons should present precisely the dual tangent bundle function associated with $P_G$. 
Indeed, using the algebraic classification of area metrics \cite{schuller2010}, it is then a simple exercise to verify that for metaclasses I--XI and XIII--XIX\new{,} the cotangent bundle function $P_G^{\#}$ defined in (\ref{areametricdualpolynomial}) satisfies at every point the defining property of the dual polynomial (\ref{dualpolynomial}). But as we will see in the illustrations at the end of the next section, area metrics of metaclasses VIII--XXIII can never give rise to viable dispersion relations. Anticipating that result, we recognize $P_G^{\#}$ as the dual polynomial to $P_G$ for all viable area metric spacetime geometries.

%\newpage
\Section{Bi-hyperbolic and energy-distinguishing dispersion relations}
\label{sec_bihyperbolicity}

{\it In this section, bi-hyperbolicity (meaning that both $P$ and $P^\#$ are hyperbolic) and the energy-distinguishing property are introduced as further conditions on the geometry. These further conditions are needed in order to provide an unambiguous notion of observers and positive energy. }\\

We saw that the dual polynomials defined by the tangent bundle function $P^\#$ in each tangent space play an essential r\^ole \new{}. In fact, one needs to restrict attention to dispersion relations that are bi-hyperbolic, meaning that both $P$ and its dual $P^\#$ are hyperbolic. This is because only then may one select one hyperbolicity cone $C^\#$ of $P^\#$ (\new{which is defined, mutatis mutandis, precisely as the hyperbolicity cones of} $P$ in section \ref{sec_hyperbolicity}) and stipulate that it contain the tangent vectors to admissible observers at a spacetime point. We will prove a non-trivial consistency result concerning the stability of \new{the so} defined observers in section \ref{sec_energycut}. \new{Physically, a} choice of $C^\#$ corresponds to a choice of time orientation of the manifold. The point is that having chosen the observer cone $C^\#$ in the tangent bundle, one can immediately see that those momenta $p$ at a point $x$ whose energy is positive from every observer's point of view, constitute a convex cone 
\begin{equation}
 (C_x^\#)^\perp = \{p \in T^*_xM \,|\, p(v)>0 \textrm{ for all } v \in C^\#\}\,.
\end{equation} 
For the Lorentzian metric case, this is simply \new{(the closure of) what has been chosen as the forward cone}, while \new{in general} the situation is more complicated; see the illustration at the end of this section.  
 If the polynomial $P$ is of the product form (\ref{reducedpolynomial}), we find that the positive energy cone is simply the sum of the positive energy cones coming from the duals of the factors of $P$ \cite{rockafellar1970}, i.e. 
\begin{equation}
\label{carpetdualsum}
 (C^\#)^{\bot}=(C_1^\#)^{\bot}+\cdots + (C_l^\#)^{\bot}\,,
\end{equation}
where the sum of two convex sets is just the set of all sums of \new{any} two elements of the two sets.
\begin{figure}[h]
\includegraphics[width=11cm,angle=0]{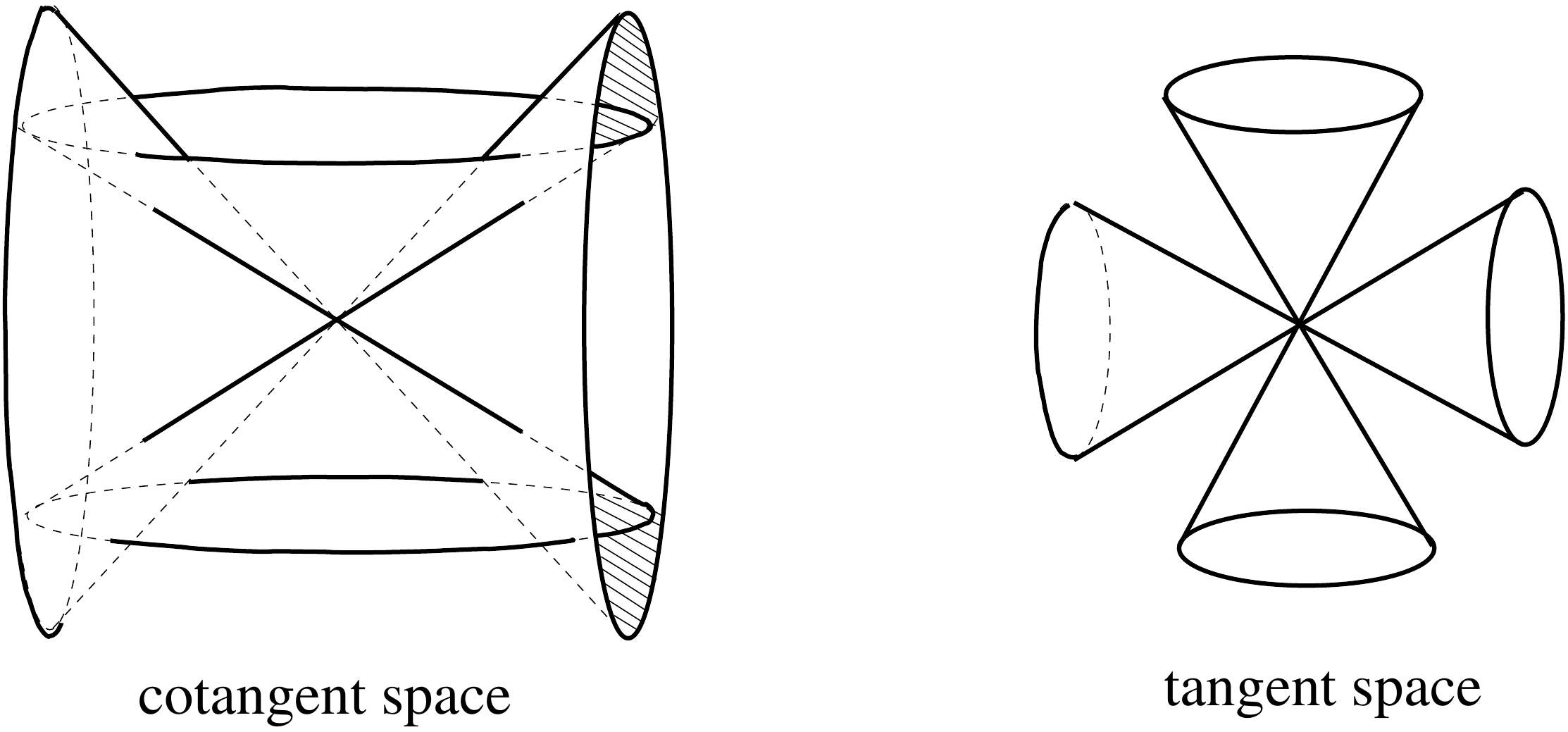}
\caption{\label{fig_bihyperbolic} Example of a hyperbolic polynomial with non-hyperbolic dual polynomial.}
\end{figure}

That hyperbolicity does not imply bi-hyperbolicity is illustrated by the counterexample in figure \ref{fig_bihyperbolic}. Bi-hyperbolicity indeed presents a rather stringent condition on dispersion relations, and thus on the underlying geometry. This is illustrated at the end of this section, first for the case of metric geometry, where bi-hyperbolicity amounts to the requirement that the metric be Lorentzian, and second for the case of area metric geometry, where a similar exclusion of algebraic classes follows directly from bi-hyperbolicity. A similar study may and needs to be conducted for the reader's favourite candidate for a spacetime geometry. In the following we focus on some conclusions that can be drawn independent of any particular geometry and which will be important for our further theoretical developments.

\begin{figure}[h]
\includegraphics[width=16cm,angle=0]{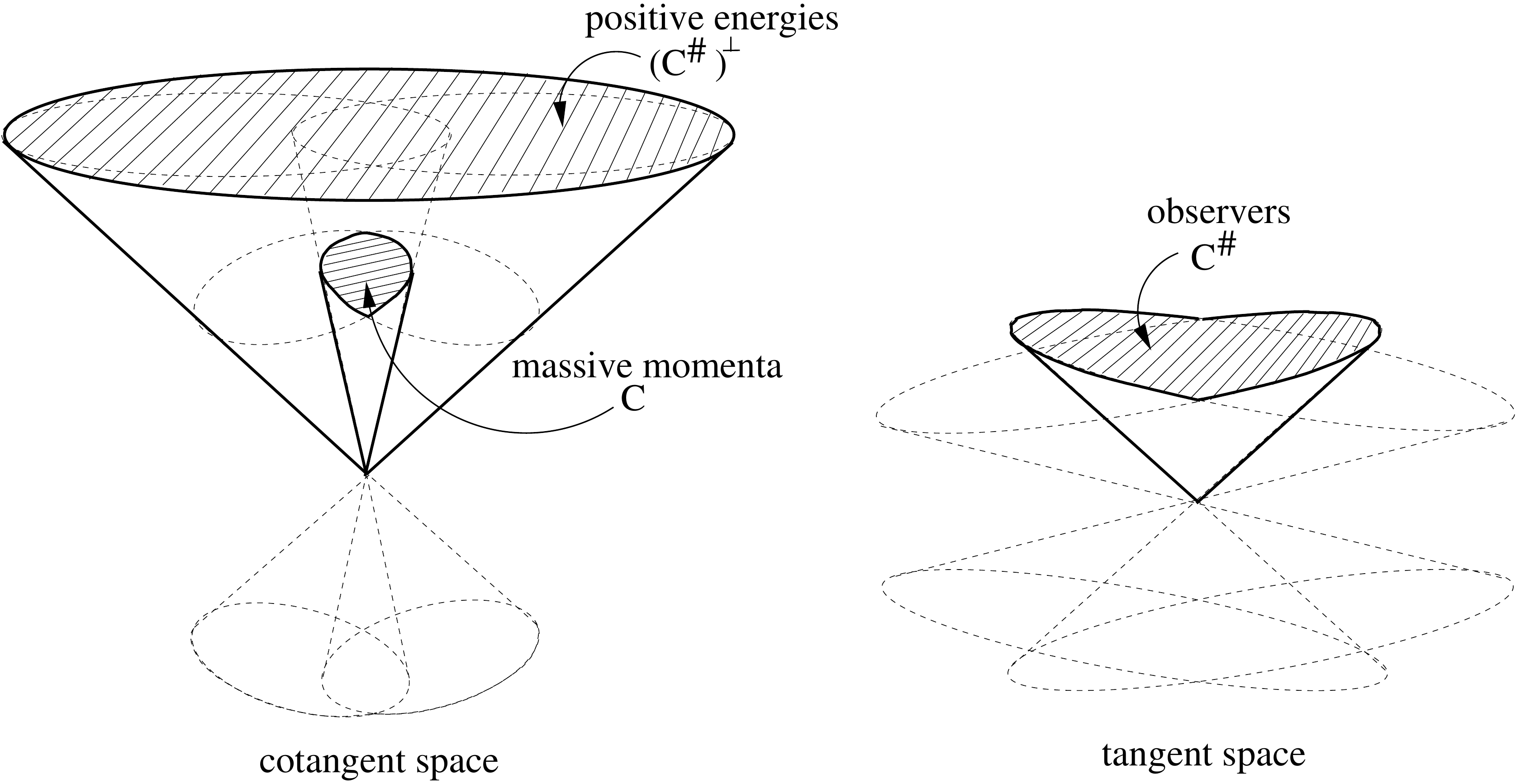}
\caption{\label{fig_allcones} An energy-distinguishing bi-hyperbolic dispersion relation.}
\end{figure}

Having guaranteed an observer independent notion of positive energy, the only \new{thing} left is to ensure that any massless momentum $q$ has either positive or negative energy. More precisely, we require the set $N$ of massless non-zero covector fields to disjunctively decompose into positive and negative energy parts
\begin{equation}
  N =N^+ \,\dot\cup \,N^-\,,
\end{equation} 
where $N^+$ is defined as the intersection of $N$ with the positive energy cone $(C^\#)^\perp$, and $N^-$ as the intersection with the negative energy cone $(-C^\#)^\perp$. We will refer to such bi-hyperbolic cotangent bundle functions $P$ as energy-distinguishing.
Figure \ref{fig_allcones} shows the vanishing sets of an energy-distinguishing bi-hyperbolic polynomial $P_x$ and its dual $P_x^\#$.

For dispersion relations that are both bi-hyperbolic and energy-distinguishing, we find that the set of massless momenta $N_x$ cannot contain any null planes in spacetime \new{dimensions} $d\ge 3$, which in turn implies that \new{} the degree of $P$ cannot be odd. This will be of importance later, and is seen as follows: First, we prove that bi-hyperbolicity of $P_x$ implies that 
\begin{equation}\label{posnegcone}
\new{\textrm{closure}(C_x^{\#\bot})\cap -\textrm{closure}(C_x^{\#\bot})=\{0\}\,.}
\end{equation} 
Let $k_0$ be such that $k_0\in \textrm{closure}(C_x^{\#\bot})$ and $k_0\in -\textrm{closure}(C_x^{\#\bot})$. It follows from the definition of the dual cone that the following inequalities are true for all $x\in C_x^\#$ : $x.k_0\ge0$ and $x.k_0\le0$. If this would be true the hyperbolicity cone $C_x^\#$ had to be a plane or a subset of a plane. That would contradict the property of $C_x^\#$ to be open. Second, suppose that the zero set $N_x$ contains a plane. From $\textrm{closure}(C_x^{\#\bot})\cap -\textrm{closure}(C_x^{\#\bot})=\{0\}$ it follows that $C_x^{\#\bot}\setminus \{0\}$ is a proper subset of a halfspace. A proper subset of a halfspace cannot contain any complete plane through the origin. Hence the \new{existence} of a null plane of $P_x$ would obstruct the energy-distinguishing property. 
Third, this \new{fact} immediately restricts us to cotangent bundle functions $P$ of even degree. For suppose $\deg P$ was odd. Then on the one hand, we would have an odd number of null sheets. On the other hand, the homogeneity of P implies that null sheets in a contangent space come in pairs, of which one partner is the point reflection of the other. Together this implies that we would have at least one null hyperplane. \\

At this point in this article, we have \new{arrived at the insight} that a physical dispersion relation for massless point particles must be given by a cotangent bundle function $P$ that is a bi-hyperbolic and energy-distinguishing reduced hyperbolic homogeneous polynomial in each fibre. These are now all the conditions on $P$ we identify in this work. The following two \new{sections} serve to show that the \new{theory extends to} massive dispersion relations and \new{allows for} all kinematical constructions one needs to provide a physical interpretation of quantities on the spacetime manifold. The final two sections are then devoted to further embellish the theory, in particular to make contact to non-covariant representations of dispersion relations and to derive a generic mechanism for a covariant energy cut-off on non-metric spacetimes.   

\subsection{Illustration: Exclusion of algebraic classes of metrics and area metrics}\enlargethispage{1cm}
The polynomial defining the massless dispersion relation on a metric background $g$ is given as $P_{g\,x}(q)=g_x^{-1}(q,q)$, as seen in section \ref{sec_covdisprelmassless}. It is easy to verify that $P_g$ is bi-hyperbolic if and only if $g$ is a metric of Lorentzian signature and that its dual is given as $P_{g\,x}^\#(v)=g_x(v,v)$. Moreover, hyperbolicity of $P_g$ also implies the energy-distinguishing property. For from the explicit definition of $C^\#$ we know that at every point $x$ and for every vector $X\in C_x^\#$ the covector $g_x(X,\cdot)\in (C_x^\#)^\perp$. Arranging for $P_{g\,x}^\#(C_x^\#)>0$ and knowing that $P_g$ is hyperbolic, it is easy to show that $g_x(\omega,v)>0$ for every vector $\omega \in \partial C_x^\#$ and $v \in C_x^\#$, which shows that for every vector  $\omega\in \partial C_x^\#$ the covector $g_x(\omega,\cdot)\in (C_x^\#)^\perp$ and
$g_x(-\omega,\cdot)\in -(C_x^\#)^\perp$. More precisely $g_x(\partial C_x^\#,\cdot)\in (C_x^\#)^\perp$ and $g_x(-\partial C_x^\#,\cdot)\in
-(C_x^\#)^\perp$. But $g_x(\partial C_x^\#,\cdot)\in (C_x^\#)^\perp$ is the image of the dual Gauss map induced from ${P_{g\,x}}^\#$ when
applied to $\partial C_x^\#$. Thus, we conclude that Lorentzian metric geometry is a bi-hyperbolic geometry of the energy-distinguishing type.

For the area metric case,  bi-hyperbolicity serves to exclude the algebraic area metric \new{metaclasses} VIII to XXIII. That result was implicitly obtained in \cite{schuller2010}, specifically  it is contained in Lemma 4.1 of that work, which asserts that there exists a plane of massless covectors for any four-dimensional area metric manifold belonging to metaclasses VIII to XXIII. But since the existence of a null plane does not allow for a bi-hyperbolic and energy-distinguishing $P$, the area metric algebraic metaclasses VIII to XXIII must be discarded as viable spacetime geometries. 

%\newpage
\Section{Covariant dispersion relations II: \quad Massive particles}\label{sec_massive}

{\it In this section, we extend the theory to massive dispersion relations and define the set of positive energy massive particles. 
Bi-hyperbolicity \new{and the energy-distinguishing property}, orginally introduced to guarantee a well-defined duality theory between massless covectors and massless vectors, \new{are} shown to also play \new{a} crucial r\^ole when discussing massive matter. In particular, \new{they ensure} reverse triangle and \new{inverse} Cauchy-Schwarz inequalities.}\\

For a bi-hyperbolic and energy-distinguishing dispersion relation, there is always a hyperbolicity cone in cotangent space that is of positive energy with respect to a chosen time orientation $C^\#$.    
For let $\tilde C_x$ be \new{some} hyperbolicity cone of $P_x$, whose boundary $\partial \tilde C_x$ we know to be a connected set of null covectors.  Now on the one hand, the complete zero set of $P_x$ is contained in $(\tilde C_x^{\#})^{\bot}\,\cup-(\tilde C_x^{\#})^{\bot}$ due to the energy-distinguishing property. On the other hand, we have \new{that (\ref{posnegcone}) holds.} Hence either $\tilde C_x$ or $-\tilde C_x$ is of positive energy. 

\begin{figure}[h]
\includegraphics[width=16cm,angle=0]{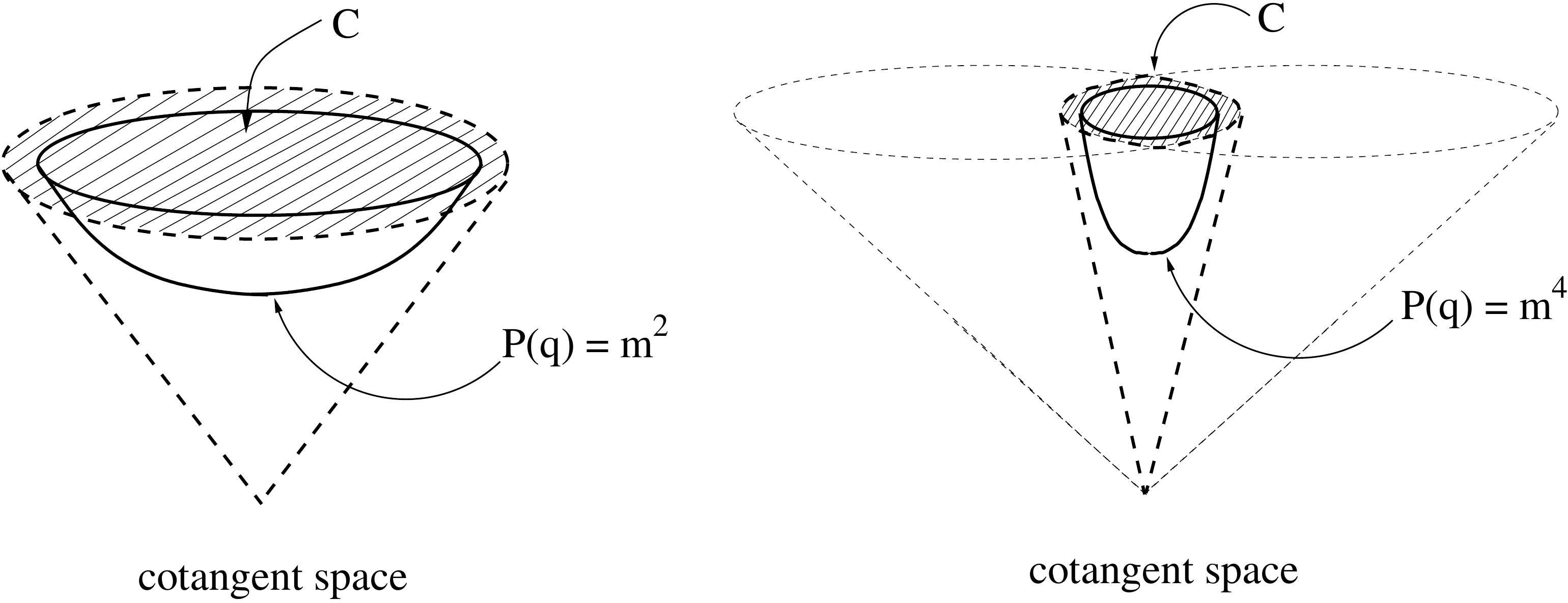}
\caption{\label{fig_massive} Mass shells defined by bi-hyperbolic energy-distinguishing cotangent bundle functions $P$.  On the left the familiar second degree Lorentzian case; on the right a fourth degree case defined by a product of two Lorentzian metrics. }
\end{figure}

The covector fields in the thus selected positive energy cone $C$ play two related r\^oles. The first r\^ole, from the point of view of the field theory, was that a hypersurface can only be an initial data surface if its normal covector field lies in $C$.
Now in order to identify the second r\^ole of the cone $C$ in relation to massive matter, first observe that within the hyperbolicity cone $C$, the sign of $P$ cannot change, so that we may arrange for $P$ to be positive on $C$ without upsetting any of the constructions made so far. \new{We will assume from now on without further comment that this choice has been made.}
But then we have for any momentum $q$ in $C_x$ at a spacetime point $x$ that
\begin{equation}\label{massivedispersion}
  P_x(q) = m^{\deg P}
\end{equation}
for some positive real number $m>0$, which we call the mass associated with the momentum $q$. It must be emphasized that the definition of mass associated to a momentum, as provided by (\ref{massivedispersion}), hinges on the choice of a particular volume \new{density} $\omega_G$ in (\ref{Pdef}). Physically this is understood from the need to convert mass densities in field theory into point masses in particle theory, which conversion requires a definition of volume. But then (\ref{massivedispersion}) represents a massive dispersion relation whose mass shells foliate the interior of $C_x$, see figure \ref{fig_massive}. An immediate physical consequence of the convexity of the cone $C_x$ is that even for modified dispersion relations, a decay
%\begin{figure}[h]
%\includegraphics[width=10cm,angle=0]{photondecay.pdf}
%\caption{\label{fig_hyperbolic} Every vector in the shaded cone, and its point mirror image, is hyperbolic}
%\end{figure}
 of a positive energy massless particle into positive energy massive particles is kinematically forbidden.

At this point we derive a further important consequence of bi-hyperbolicity and the energy-distinguishing property, namely that together they imply completeness of the polynomials $P_x$ defined by $P$ in each cotangent space; \new{in the terminology of  \cite{garding1959},} a hyperbolic polynomial $P_x$ is called complete if the lineality space 
\begin{equation}\label{lineality}
  L(P) = \{a \in T_x^*M \,|\, \textrm{for all } y\in T_x^*M \textrm{ and } \lambda \in \mathbb{R}: P(y+\lambda a)=P(y)\}
\end{equation}
only contains the zero covector. In other words, in order to be complete\new{,} $P$ must depend on all covector components in any chosen basis. Geometrically, completeness can be read off from the closure of the hyperbolicity cones since \cite{bauschke2001} it is equivalent to  
  \begin{equation}
  \textrm{closure}(C(P_x,h)) \,\,\cap\,\, \textrm{closure}(C(P_x,-h)) = \{0 \}\,.
\end{equation}  

That completeness is already implied by the energy-distinguishing condition can be easily seen from this. For picking up the argument given at the start of this section, we know that
\begin{equation}
\new{\textrm{closure}(C_x^\#{}^\perp)\cap -\textrm{closure}(C^\#_x{}^\perp) \supseteq \textrm{closure}(C_x) \cap - \textrm{closure}(C_x)\,.}
\end{equation}
 Thus if the right hand side differs from $\{0\}$ (meaning that $P$ \new{is} incomplete), the left hand side will contain non-zero covectors, too (showing that $P$ is not energy-distinguishing). Because of the inclusion, this \new{} only holds in this direction. We conclude that the energy-distinguishing property already implies completeness.

There are three principal reasons why it is so important that completeness holds. 
First, completeness will play a crucial r\^ole in ensuring, as we will see in the next section, that there is a \new{} well-defined duality theory associating massive covectors with their vector counterparts. Thus remarkably, bi-hyperbolicity \new{and the energy-distinguishing property}, originally conceived in the context of massless dispersion relations, also \new{take} care of \new{this in the massive case}, via completeness.  
Second, since we arranged for $P_x$ to be positive everywhere on $C_x$ for the massive dispersion relation to make sense, we have the reverse triangle inequality
\begin{equation}
  P_x^{1/\deg P}(k_1 + k_2) \geq P_x^{1/\deg P}(k_1) + P_x^{1/\deg P}(k_2)  
\end{equation}  
for all $k_1$ and $k_2$ in the same hyperbolicity cone $C_x$. Equality holds if and only if $k_1$ and $k_2$ are all proportional. Physically, the reverse triangle inequality generalizes a familiar result from Lorentzian geometry to any viable dispersion relation in our sense, namely that the decay of a massive particle generically gives rise to a mass defect.
 Third, defining the tensor
\begin{equation}
\label{polarizatidP}
   P_x(k_1, \dots, k_{\deg P}) = \frac{1}{(\deg P)!} \prod_{J=1}^{\deg P} \left(\sum_{i=1}^{\dim V} (k_J)_i \frac{\partial}{\partial k_i}\right) P_x(k)\,,  
\end{equation}
as the totally symmetric polarization of the polynomial $P_x$, 
we can also formulate a reverse Cauchy-Schwarz inequality
\begin{equation}
\label{reverse_Cauchy-Schwarz}
  P_x(k_1, \dots, k_{\deg P}) \geq P_x(k_1)^{1/\deg P}  \cdots P_x(k_{\deg P})^{1/\deg P} 
\end{equation}
for all $k_1, \dots, k_{\deg P}$ in the same hyperbolicity cone $C_x$. 
Similar to the reverse triangle inequality above, equality holds for the reverse Cauchy-Schwarz inequalities if and only if all arguments $k_i$ are proportional to each other. 

\Section{Vector duals of massive momenta: Legendre map}\label{sec_legendre}

{\it For a bi-hyperbolic and energy-distinguishing $P$, we introduce the action for free massive point particles. In contrast to the polynomial tangent space geometry $P^\#$ seen by massless particles, the dual geometry seen by massive particles \new{turns out to be} encoded in a generically non-polynomial tangent bundle function $P^*$. This largely obstructs attempts to devise a non-trivial Finslerian or Lagrangian tangent bundle geometry that describes massive and massless point particles simultaneously.}\\

 We wish to associate vector duals \new{} with \new{} massive momenta (having done so for massless momenta in chapter \ref{sec_masslessduals}), and to this end we employ the Helmholtz action
\begin{equation}\label{Helmmassive}
  I[x,q,\lambda] = \int d\tau \left[q_a \dot x^a - \lambda m \ln P(x,\frac{q}{m})\right]\,,
\end{equation} 
which describes particles that are free due to the form of the first term, and massive since the massive dispersion relation $P(x,q)=m^{\deg P}$ is enforced through variation with respect to $\lambda$. The particular form of the Lagrange multiplier term here has been chosen for the technical reason of having available the theory of Legendre duals on the open convex cones $C_x$, see \cite{rockafellar1970}. More precisely, the so-called barrier function, 
\begin{equation}
\label{barrierfunction}
  f_x: C_x \to \mathbb{R}\,,\qquad f_x(q) = -\frac{1}{\deg P}\ln P_x(q)\,,
\end{equation}
which we employed in the massive particle action above, 
is firstly guaranteed to be strictly convex, i.e., for each $\lambda\in [0,1]$ we have $f_x((1-\lambda)v + \lambda w) < (1-\lambda)f_x(v) + \lambda f_x(w)$ for all $v,w$ in the hyperbolicity cone $C_x$, due to the completeness of $P$ \cite{bauschke2001}, which in turn is guaranteed by the energy-distinguishing property, as we saw in the previous chapter;
 secondly, near the boundary of the convex set, it  behaves such that for all $q\in C_x$ and $b\in \partial C_x$
\begin{equation}
  \lim_{\lambda\to 0^+}(D_{q-b}f_x)(b+\lambda(q-b)) = 0\,,
\end{equation}
which property is known as essential smoothness in convex analysis. 
The important point is that \new{} strict convexity and essential smoothness \new{together} ensure that the barrier function $f_x$ induces an invertible Legendre map
\begin{equation} \label{Ldef}
  L_{x}: C_x \to L_{x}(C_x)\,,\qquad q \mapsto -(D f_x)(q)
\end{equation}
and a Legendre dual function
\begin{equation}\label{oneeqn}
  f_x^L: L_x(C_x) \to \mathbb{R}, \quad f_x^L(v) = -L_{x}^{-1}(v) v - f_x(L_{x}^{-1}(v))\,\new{,} 
\end{equation}
which can be shown, ultimately by virtue of the above conditions, to be an again strictly convex and essentially smooth function on the open convex set $L_x(C_x)$. Note that the two minus signs in (\ref{oneeqn}) are correct, and due to our sign conventions. In fact, the inverse Legendre map is the Legendre map of the \new{Legendre dual function $f^L$}:
\begin{equation}
  -Df_x^L = L_x^{-1}(v) + DL_x^{-1}(v)v + DL_x^{-1}(v) Df_x(L_x^{-1}(v)) = L_x^{-1}(v)\,.
\end{equation}
In other words, the Legendre dual of the Legendre dual $(L_x(C_x),f_x^L)$ of $(C_x,f_x)$ is again $(C_x,f_x)$, see theorem 26.5 of \cite{rockafellar1970}.
\begin{figure}[h]
\includegraphics[width=15cm,angle=0]{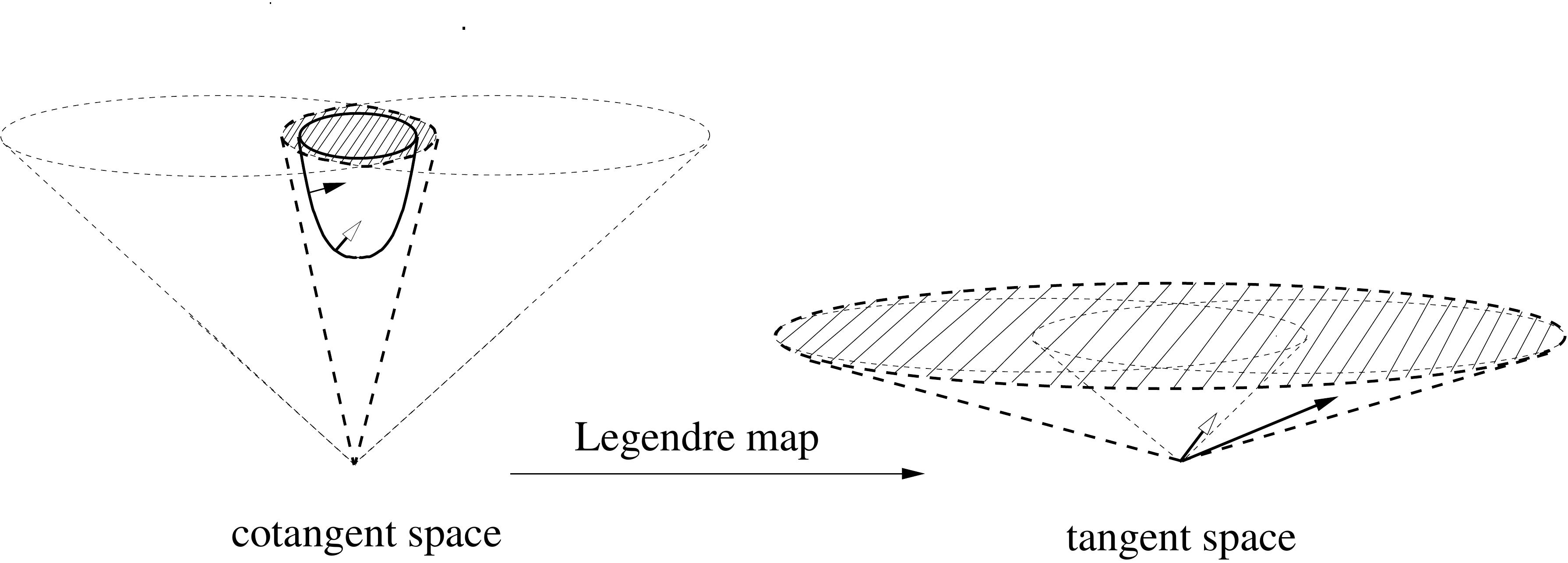}
\caption{\label{fig_legendre} Mass shell and Legendre map of massive momenta to tangent space.}
\end{figure}

The existence of this Legendre theory now enables us to eliminate the $q$ and $\lambda$ degrees of freedom, in order to obtain an equivalent particle action $I[x]$ in terms of the particle trajectory $x$. In the process, we will identify the definition of proper time that renders the law of free particle motion simple. Variation \new{of the action (\ref{Helmmassive})} with respect to $q$ yields 
$\dot x = (\lambda  \deg P) L_x(q/m)$, which we know may be inverted to yield 
\begin{equation}
   q = m  L_x^{-1}(\dot x/(\lambda\deg P))\,.
\end{equation}
It is now obvious why it was convenient to encode the dispersion relation by a Lagrange multiplier term involving the barrier function (\ref{barrierfunction}); while many other ways to enforce the very same dispersion relation of course do exist, the latter allows to make use of the above theory of Legendre transformations in a straightforward manner. Using the thus obtained relation and the definitions of the barrier function and the Legendre dual to eliminate $q$, one obtains the equivalent action 
\begin{equation}\label{intermediate}
  I[x,\lambda] \new{= - m \deg P \int d\tau \, \lambda f^L(\dot x/(\lambda \deg P))} = -m \deg P\int d\tau \left[\lambda f_x^L(\dot x) + \lambda \ln(\lambda\deg P)\right]\,,
\end{equation}
where for the second equality we used the easily verified scaling property $f^L(\alpha \dot x) = f^L(\dot x) - \ln \alpha$. From variation \new{of the action (\ref{intermediate})} with respect to $\lambda$\new{,} we then learn that 
\begin{equation}
  f^L(\dot x) + \ln(\lambda \deg P) + 1 = 0\,.
\end{equation}
Using this twice, we have $\lambda  f_x^L(\dot x) +\lambda \ln (\lambda \deg P))= -\lambda = - \exp(-f_x^L(\dot x)-1)/\deg P$. Noting that because of $\dot x\in L_x(C_x)$ we also have $L^{-1}(x,\new{\dot x})(\dot x) = 1$ and thus $f_x^L(\dot x)=-1-f_x(L^{-1}(\dot x))$,
and defining the tangent bundle function
\begin{equation}
  P_x^*: L_x(C_x) \to \mathbb{R}\,, \qquad P_x^*(v) = P_x(L_x^{-1}(v))^{-1}\,,
\end{equation} 
\new{we eliminate $\lambda$ in (\ref{intermediate}) and finally arrive at} the equivalent action
\begin{equation}
\label{finsleraction}
  I[x] =m \int d\tau P^*(x,\dot x)^{1/\deg P}\new{}
\end{equation}
for a free point particle of positive mass $m$. 
 While the tangent bundle function $P^*$ is generically non-polynomial, it is elementary to see that it is homogeneous of degree $\deg P$, and for later reference we also display the useful relation
\begin{equation}\label{PstarandL}
  L_x^{-1}(v) = \frac{1}{\deg P} \frac{D P_x^*(x,v)}{P_x^*(x,v)}\,.
\end{equation} 
The action (\ref{finsleraction}) is reparametrization invariant, as it should be. However, parametrizations for which $P(x,L^{-1}(x,\dot x)) = 1$ along the curve are distinguished since they yield the simple relation 
\begin{equation}\label{simplemomentum}
  \dot x = L_x(q/m)
\end{equation}
between the free massive particle velocity $\dot x$ and the particle momentum $q$ everywhere along the trajectory $x$. As usual, we choose such clocks and call the time they show proper time.
 Thus we have established the physical meaning of the Legendre map, and may thus justifiedly call the open convex cone $L_x(C_x)$ the cone of massive particle velocities, and the function $P^*$ the massive dual of $P$, which indeed encodes the tangent bundle geometry seen by massive particles.

Reassuringly, we can now prove that the observer cone lies in the massive dual,  $C_x^\# \subseteq L_x(C_x)$. Thus one may think of observers as massive, as usual. The converse, however, does not hold, since the inclusion is generically proper. Since this statement, in slightly refined form, will be of central importance again in chapter \ref{sec_energycut}, we will formulate it by way of two lemmas:\\

\noindent{\bf Third Lemma.} For any reduced hyperbolic homogeneous cotangent bundle function $P$ we have $L_x(C_x)=\textrm{interior}(C_x^{\bot})$.\\

{\it Proof.} Since by assumption $P_x$ is \new{reduced, hyperbolic and homogeneous,} we get from the First and the Second Lemma in chapter \ref{sec_hyperbolicity} the statement: for all $p\in T_x^*M \setminus \textrm{closure}(C_x)$ there exists an $r\in\partial C_x$ such that $p.D P_x(r)<0$. Since $p.D P_x(q)$ is a continuous function of $q$, we conclude that for all $p\in T_x^*M \setminus \textrm{closure}(C_x)$ there exists an $q\in C_x$ such that $p.D P_x(q)<0$. That implies that the set $L_x(C_x)^{\bot}$ is a subset of $\textrm{closure}(C_x)\setminus\{0\}$. Since $L_x(C_x)$ is convex, we get $L_x(C_x)\supseteq (\textrm{closure}(C_x)\setminus\{0\})^{\bot}=\textrm{interior}(C_x^{\bot})$. Furthermore, we know that $L_x(C_x)\subseteq C_x^{\bot}$. Since $L_x(C_x)$ is open it follows that $L_x(C_x)= \textrm{interior}(C_x^{\bot})$.\\

\noindent{\bf Fourth Lemma.} For any bi-hyperbolic and energy-distinguishing cotangent bundle function $P$, we have $C_x^\# \subseteq \textrm{interior}(C_x^{\bot})$.\\

{\it Proof. } From chapter \ref{sec_massive} we know that there exists a hyperbolicity cone $C_x$ of $P_x$ that lies completely in $(C_x^\#)^{\bot}$. From $(C_x^\#)^{\bot}\supseteq C_x$ and the fact that $C_x^\#$ is open, we conclude that $C_x^\#\subseteq \textrm{interior}(C_x^{\bot})$.\\

Comparing these results with those of chapter \ref{sec_masslessduals}, we see that there is a fundamental difference between the ways in which null covectors on the one hand, and massive covectors on the other hand, are mapped to the respective velocities on tangent space. In the null case, the Gauss maps $[DP_x]$ and $[DP_x^\#]$ associate massless particle momenta with the respective null velocities, up to an undetermined real factor. In the massive case, in contrast, the Legendre map $L_x$ and its inverse $L_x^{-1}$ afford the same for massive particle momenta and velocities. \new{As a consequence,} the dual geometries seen on the tangent bundle by massless and massive particles differ. For the former, the Gauss dual $P^\#$ is the relevant structure, and for the latter the Legendre dual $P^*$. We wish to emphasize again that while $P^\#$ is polynomial in its fibre argument, $P^*$ generically \new{is} not. Indeed, to explicitly find the inverse Legendre map $L^{-1}$, and thus $P^*$, can be very hard in concrete applications, although its existence and uniqueness are guaranteed. Also in this sense, the tangent bundle geometry $(TM,P^\#,P^*)$ is considerably less straightforward than the cotangent bundle geometry $(T^*M,P)$ it dualizes. 
This explains \new{to some extent the difficulties} noticed by Skakala and Visser in \cite{skakala2009},\cite{skakala2010} to identify a single Finsler-type tangent bundle geometry: generically there simply is no such geometry on tangent space that could give rise, dually, to a bi-hyperbolic energy-distinguishing dispersion relation. The case of a Lorentzian geometry presents one notable exception.

 On the positive side, on the cotangent bundle, any bi-hyperbolic and energy-distinguishing reduced homogeneously polynomial geometry provided by $P$ provides a perfectly fine spacetime geometry as far as point particle theory is concerned. And if one wishes to consider the coupling of fields, one needs to couple to an underlying tensorial geometry $G$ that gives rise, by the very same field equations, to the cotangent bundle function $P$ at hand, as discussed in chapter \ref{sec_covdisprelmassless}.  Again, the metric case is degenerate, since there\new{} one does not recognize the difference between the four different r\^oles played by the metric: \new{the} inverse metric plays the r\^ole of the fundamental spacetime structure to which fields couple, as well as the r\^ole of defining the (structurally very different) cotangent bundle function $P$, while the metric plays the r\^ole of both the dual $P^\#$ as well as the tangent bundle function $P^*$, which define the tangent space geometries seen by massless and massive particles, respectively. All these different structures are, strictly conceptually speaking, of course already at play in the familiar metric case, but display their different nature explicitly only in the general case.

%\newpage
%\Section{$E(\vec{p})$ representation of the covariant dispersion relation}
%\Section{Non-covariant dispersion relations and inertial laboratories} 
\Section{$E(\vec{p})$ form of dispersion relations and inertial laboratories} 
%\Section{$E(\vec{p})$ representation and freely falling non-rotating laboratories}
\label{sec_E(p)}

{\it The decidedly covariant discussion of dispersion relations, which we adopted in this paper, is required even if one ultimately prefers to represent the dispersion relation in form of a function $E(\vec{p})$\new{, which expresses} the energy of a particle in terms of its spatial momentum. This is because the therefore needed temporal-spatial split of a covariant particle momentum itself is governed by the covariant version of the dispersion relation.  Pushing the theory of observers and frames further, we identify \new{a} generically non-linear parallel transport induced by a bi-hyperbolic and energy-distinguishing dispersion relation and thus succeed in defining inertial laboratories.}\\

\begin{figure}[h]
\includegraphics[width=16cm,angle=0]{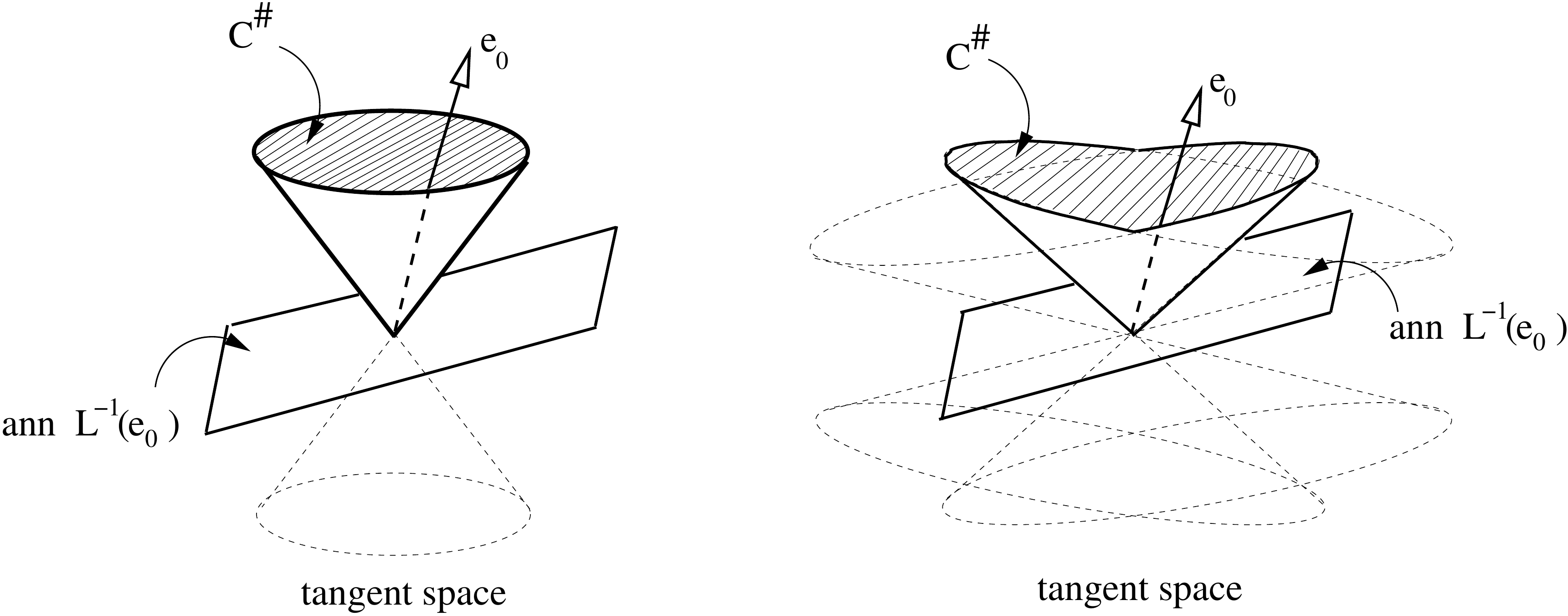}
\caption{\label{fig_observerframe} Purely spatial directions with respect to $e_0$ are those annihilated by $L^{-1}(e_0)$. For the Lorentzian metric case on the left, this is coincides with the space of vectors $g$-orthogonal to $e_0$.}

\end{figure}

Converting our covariant dispersion relations for massive and massless matter into non-covariant dispersion relations is conceptually and mathematically straightforward. 
For from the point of view of an observer carrying a clock that shows proper time (who is thus formally described by a $P^*$-unit vector $e_0$ in the observer cone $C^\#$, see the discussion following equation (\ref{simplemomentum})), any spacetime momentum $p$ can be uniquely decomposed as
\begin{equation}\label{split}
  p = E\, L^{-1}(e_0) + \vec{p}\,,
\end{equation}  
namely into an energy $E$ and a purely spatial momentum $\vec{p}$ satisfying $\vec{p}(e_0)=0$. For a visualization, see the corresponding tangent space split in figure \ref{fig_observerframe}. 
Employing such a particular observer-dependent split, one may solve the covariant dispersion relation
\begin{equation}
\label{covariantdispersionrelation}
  P(x, E \, L^{-1}(e_0) + \vec{p}) = m^{\deg P}
\end{equation}
for the energy $E$ in terms of the spatial momentum $\vec{p}$, and thus obtain an observer-dependent, non-covariant dispersion relation $E=E(\vec{p})$. 
Keeping in mind that the latter depends in two ways on the cotangent bundle function $P$, namely indirectly through the temporal-spatial split (\ref{split}) imposed by it and directly through the dispersion relation (\ref{covariantdispersionrelation}), this non-covariant version can be useful since it more directly relates to measurable quantities. However, due to Galois theory, we know that the energy will not even be an analytic expression in terms of the spatial momentum unless $\deg P \leq 4$, and not polynomial in any case. The crucial properties of bi-hyperbolicity and the energy-distinguishing property are even more hidden in the non-covariant formulation. This is of course the key reason for having dealt exclusively with a strictly covariant treatment of dispersion relations for all formal developments throughout this article. 

Conversely, the conversion of a non-covariant dispersion relation into a covariant one will be prohibitively difficult in most cases. This is essentially due to the fact that given a relation $E=E(\vec{p})$, the construction of a spacetime momentum $p$ from the $E$ and $\vec{p}$, and indeed their physical meaning, is not directly possible without the cotangent bundle function $P$. We feel that this is often not considered where modified dispersion relations are proposed. Sometimes recourse to an `anyway' underlying spacetime metric is made, but it is hard to see how this would be consistent with the stipulation of a modified dispersion relation, due to the above double r\^ole played by the cotangent bundle function $P$. 

In the above discussion, it was sufficient to identify the purely spatial momentum associated with the unit timelike direction of a particular observer worldline. It is often useful to go further and to consider freely falling non-rotating observer frames. This is needed, for instance, if one wishes to determine the electric and magnetic field strengths seen by such an observer for a given electromagnetic field strength two-form $F$.  But the definition of non-rotating frames requires to establish a meaningful parallel transport, and we will now see how the latter arises from a general dispersion relation. Since we saw in seection \ref{sec_legendre} that observers are necessarily massive, their free motion is governed by an action functional 

\begin{equation}
  S[x] = \int d\tau P^*(x,\dot x)^{1/\deg P}\,,
\end{equation} 
which we know to represent the trajectories of point particles of non-zero mass. Using the reparametrization invariance to set $P^*(x,\dot x)=1$ along the curve, it is straightforward to derive the equations of motion and, using in the following standard techniques of Finsler geometry \cite{shen2001}, to cast them into the form 
\begin{equation}\label{massivegeodesic}
  \ddot x^a + \Gamma^a (x,\dot x) = 0
\end{equation}
with the geodesic spray coefficients
\begin{equation}\label{Gammas}
  \Gamma^a{}(x,v) = \frac{1}{2} g_{(x,v)}^{am}\left(\frac{\partial g_{(x,v)\, mc}}{\partial x^b} + \frac{\partial g_{(x,v)\, bm}}{\partial x^c} - \frac{\partial g_{(x,v)\,bc}}{\partial x^m}\right) v^b v^c\,.
\end{equation}
These in turn are constructed from the tangent space metrics \new{$g_{e_0}$ defined by} 
\begin{equation}\label{finslerPstar}
  g_{(x,e_0)}(u,v) = \frac{1}{2} \left.\frac{\partial^2 P^*(x,e_0 + s u + t v)^{2/\deg P}}{\partial s \partial t} \right|_{s=t=0}\,,
\end{equation}
whose inverses appearing in the expression (\ref{Gammas}) are guaranteed to exist from the completeness of the cotangent bundle function
 $P$.  \new{Indeed, for $e_0=L(\epsilon^0)$, an explicit expression for the metric (\ref{finslerPstar}) in terms of $f^L$ is given by 
\begin{eqnarray}
\label{finslermetric}
g_{(x,e_0)\,ab}={P^*_x}^{2/degP}(e_0)\left(- (DDf_x^L(e_0))_{ab}+2{L^{-1}_x}_{a}(e_0){L^{-1}_x}_{b}(e_0)\right),
\end{eqnarray}
and for its inverse in terms of $f$ by
\begin{eqnarray}
\label{inversefinslermetric}
g_{(x,\epsilon^0)}^{ab}={P_x}^{2/degP}(\epsilon^0)\left(-(D Df_x(\epsilon^0))^{ab}+2 {L_{x}}^{a}(\epsilon^0){L_x}^{b}(\epsilon^0)\right) ,
%&=&\frac{1}{degP} P_x(\epsilon^0)^{(2/degP)-1} (DD P_x(\epsilon^0))^{ab}+\\ \nonumber
%&&\frac{1}{degP}\left(\frac{2}{degP}-1\right)P_x(\epsilon^0)^{(2/degP)-2}(DP_x(\epsilon^0))^a(DP_x(\epsilon^0))^b\,.
\end{eqnarray}
where $(DDf_x(\epsilon^0))^{ab} (DDf_x^L(L(\epsilon^0)))_{bc}=\delta^a_c$. Remarkably, the Finsler metric (\ref{finslermetric}) is automatically Lorentzian. We show this as follows. Consider a cotangent frame $\epsilon^a$ with
 $\epsilon^\alpha(L(\epsilon^0))=0$ for all $\alpha=1, \dots, \textrm{dim} M-1$, then from expression (\ref{inversefinslermetric}) it follows that
\begin{eqnarray}
g_{(x,\epsilon^0)}^{ab}\epsilon^0_a \epsilon^0_b&=&{P_x}^{2/degP}(\epsilon^0)>0,\label{nullnullcomp}\\  
g_{(x,\epsilon^0)}^{ab}\epsilon^0_a \epsilon^{\alpha}_b&=&0\label{nullspatialcomp}.
\end{eqnarray}
But since any covector $\vec p$ on the spatial hyperplane defined by $L_x(\epsilon^0)$ can be written as $\vec p=p_\alpha \epsilon^\alpha$, we have
\begin{equation}
g_{(x,\epsilon^0)}^{ab}p_\alpha\epsilon^{\alpha}_a p_\beta\epsilon^{\beta}_b=-{P}_x^{2/degP}(\epsilon^0)(D Df_x(\epsilon^0))^{ab}p_\alpha\epsilon^{\alpha}_a p_\beta\epsilon^{\beta}_b<0,
\end{equation}
where the last inequality follows from the positive definiteness of the Hessian of $f$ (see theorem 4.2 and remark 4.3 of \cite{bauschke2001}). 
Thus we conclude that the metric (\ref{inversefinslermetric}) and hence its inverse (\ref{finslermetric}) are Lorentzian.} 

The metric (\ref{finslermetric}) and its inverse (\ref{inversefinslermetric}) will be seen to provide a normalization for local frames which is preserved along free observer worldlines. The form of equation (\ref{massivegeodesic}) indeed suggests to identify a parallel transport on the manifold $M$ which, on the one hand, allows to recast the geodesic equation in the form of an autoparallel equation, and on the other hand, provides us with the means to define parallel transport also for purely spatial vectors. To this end, it is known to be convenient to define the derivative operators
\begin{equation}
  \delta_i = \frac{\partial}{\partial x^i} - \Gamma^j{}_i(x,v) \frac{\partial}{\partial v^j}\,,\qquad \textrm{ where } \quad \Gamma^i{}_j(x,v):= \frac{\partial \Gamma^i(x,v)}{\partial v^j}\,,
\end{equation}
since now one can define, in full formal analogy to the Levi-Civita connection in metric geometry, the Chern-Rund connection coefficients
\begin{equation}\label{ChernRund}
  \Gamma^i{}_{jk}(u,v) = \frac{1}{2} g_{(x,v)}^{is} \left(\delta_j g_{(x,v)\,sk} + \delta_k g_{(x,v)\,js} - \delta_s g_{(x,v)\,ik}\right)\,.
\end{equation}
These transform, due to the use of the $\delta_i$ operators, precisely as a linear connection would under a change of coordinates $x=x(\tilde x)$. It is then straightforward to see that for any vector $w\in L(C)$ and vector field $u$ on $M$, one may define a new vector field with components
\begin{equation}
  (\nabla_w u)^i = w^a \partial_a u^i + \Gamma(x,w)^i{}_{jk}  w^j u^k\,.
\end{equation}
Clearly, $\nabla_w$ acts as a derivation on vector fiels, namely $\nabla_{w}(u + v) = \nabla_{w} u + \nabla_{w} v$ and $\nabla_{w}(f u) = (w f) u + f \nabla_{w} u$ for any function $f$ and vector fields $u, v$. Thus $\nabla_w$ may be consistently extended to act on arbitrary tensor fields $S, T$ on $M$ by imposing the Leibniz rule 
\begin{equation}
\nabla_{w}(S \otimes T) = (\nabla_{w} S)\otimes T + S \otimes (\nabla_{w} T)
\end{equation}
for arbitrary tensor fields $T$ and $S$. The derivation $\nabla_w$ is not linear in its directional argument $w$, though, and thus amounts to what is often called a non-linear connection in the literature. Nevertheless, the non-linear covariant derivative $\nabla$ achieves the desired reformulation of the geodesic equation (\ref{massivegeodesic}) as the autoparallel equation
\begin{equation}
 \label{autoparallel}
  \new{\nabla_{\dot x}\, \dot x = 0 \,.}
\end{equation}

The non-linear connection $\nabla$ provides sufficient structure for the discussion of freely falling non-rotating frames. The key technical observation is that for a frame field $e_0, \dots, e_{d-1}$ that is parallely transported along the first frame vector $e_0$,
\begin{equation}\label{freelynonrot}
  \nabla_{e_0} \,e_a = 0\,,
\end{equation}
we have the conservation equation
\begin{equation}
  e_0 \left(g_{e_0}(e_a,e_b)\right) = 0\,.
\end{equation}
This means in particular that any normalization imposed on spacetime frames \new{by virtue of the metric (\ref{finslermetric})} is preserved along the worldline of a freely falling observer. In turn, (\ref{freelynonrot}) establishes a consistent notion of \new{freely falling and non-rotating observer frames, and thus inertial laboratories.}

%\newpage
\Section{High-energy signature of modified dispersion relations}
\label{sec_energycut}

{\it  \new{} The high-energy behaviour of massive matter changes not only quantitatively, but also qualitatively, if the standard relativistic dispersion relation is modified.  In particular, \new{there is a} maximum energy which a massive particle can have without radiating off, sooner or later, a massless particle. \new{More precisely, one finds that massive} matter cannot not radiate \new{off massless particles} if and only if observers can ride on it. For the standard relativistic dispersion relation, this is the familiar result that such radiation does not take place at all. For a modified dispersion relation, this reveals a covariant mechanism for an effective high energy cut-off.}\\

\begin{figure}[h]
\includegraphics[width=15cm,angle=0]{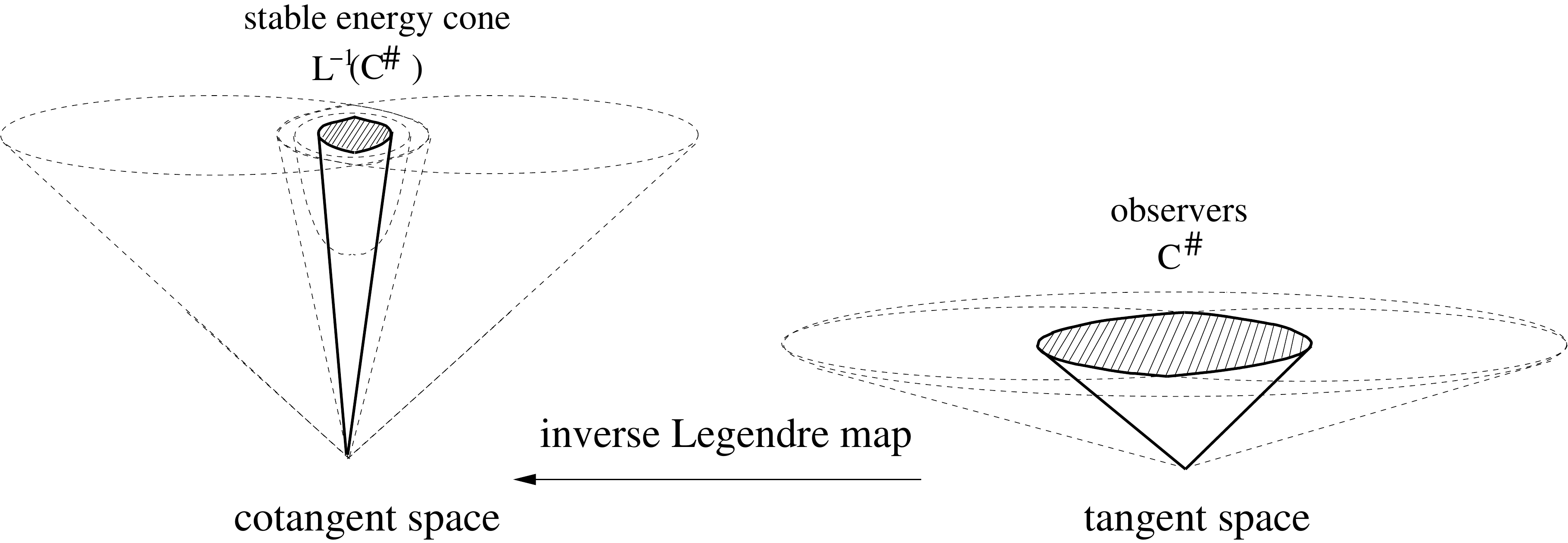}
\caption{\label{fig_stability}Stability cone: if and only if an observer can ride on a particle, the particle cannot lose energy by a vacuum Cerenkov process}
\end{figure}

Consider a process where a positive energy massive particle \new{of momentum $p$} radiates off a positive energy massless particle. 
%\begin{figure}[h]
%\includegraphics[width=12cm,angle=0]{vertex.pdf}
%\caption{\label{fig_vertex} Vacuum Cerenkov type processes.}
%\end{figure}
We will now show that due to energy-momentum conservation, such a process is kinematically forbidden if and only if $p$ lies in the stability cone 
\begin{equation}\label{stabco}
  L_x^{-1}(C_x^\#)\,,
\end{equation}
which in turn always lies entirely within the cone $C_x$ of massive momenta with positive energy. 
\new{For the proof of these assertions, see further below; for} an illustration, see figure \ref{fig_stability}. 

Specializing to the familiar case of the standard dispersion relation on a Lorentzian manifold, one of course obtains that $L_x^{-1}(C_x^\#) = C_x$; in other words, there is no Cerenkov radiation in vacuo. But for a modified dispersion relation where $L_x^{-1}(C_x^\#) \subsetneq C_x$, there is a clear covariant mechanism for a dynamic energy cut-off: if a massive particle is made so energetic that its momentum leaves $L_x^{-1}(C_x^\#)$, it may sooner or later radiate off a massless particle, or several ones, until its momentum lies within the stability cone. Reassuringly, observers' energies are obviously always below the energy cut-off.\\

Now we turn to the proof of the assertion that the stability cone (\ref{stabco}) contains precisely the momenta of those \new{massive} particles that cannot radiate off a massless particle in vacuo. To this end we will need to employ all four lemmas previously proven in this article:  First of all we get from the Third and Fourth Lemma that every observer corresponds to a massive momentum, $C_x^\# \subseteq L_x(C_x)= \textrm{interior}(C_x^{\bot})$, so that $L_x^{-1}(C_x^\#)$ is well defined and always lies within $C_x$.
It is now easy to see that a massive particle of mass $m$ and positive energy momentum $p$ may only radiate off a positive energy massless particle if there exists a positive energy massless momentum $r \in N_x^+$ such that $r(L_x(p))>0$. For consider the function
\begin{equation}
  u(\lambda) := -\ln P_x\left(\frac{p-\lambda r}{m}\right)\,.
\end{equation}
Since for any positive $\lambda$, the covector $-\lambda r \in - (C_x^\#)^\perp$ lies in some half-space of the cotangent bundle, while $p\in C_x \subset (C_x^\#)^\perp$ lies in the corresponding other half, we conclude that for some $\lambda_0>1$ the line $p-\lambda r$ will necessarily intersect the boundary of $C_x$, so that 
$\lim_{\lambda\to\lambda_0} u(\lambda_0) = +\infty$. 
Further, from theorem 4.2 and remark 4.3 of \cite{bauschke2001}, we know that for a complete hyperbolic $P_x$ the Hessian of the barrier function $- \ln P_x$ is positive definite. Hence, we find that $u''(\lambda)>0$ everywhere on its domain. 
 Now first assume that the massive particle of momentum $p$ decays into a massive particle of the same mass and of momentum $p-r$ and a massless particle of momentum $r$, thus respecting energy-momentum conservation. Then we have from the equality of masses for the ingoing and outgoing massive particles that $u(0)=u(1)=0$. But because $u''(\lambda)> 0$, the only way for the analytic function $u$ to take the same finite values at $\lambda=0$ and $\lambda=1$ while tending to $+\infty$ at some $\lambda_0>1$ is to have $0>u'(0) = -r(L_x(p))$. Conversely, assume that $r(L_x(p))>0$ for some $r\in N_x^+$. Then $u'(0)<0$ and we conclude by the mean value theorem that there must be a (because of $u''(\lambda)> 0$ unique) $\lambda_1$ with $0<\lambda_1<\lambda_0$ such that $u(\lambda_1)=0$, i.e., there is an outgoing particle of the same mass such that the process occurs.     
In summary, a massive particle of momentum $p$ can radiate off a positive energy massless particle if and only if there exists an $r\in N_x^+$ such that $r(L_x(p))>0$.

Now on the one hand, we have that $p \not\in L_x^{-1}(C_x^\#)$ if $r(L_x(p))<0$ for some $r\in N_x^+$. For then $r$ lies certainly in $(C_x^\#)^\perp$, and thus $r(L_x(p))>0$ for all $p\in L_x(C_x^\#)$. On the other hand, if $p\not\in L_x^{-1}(C_x^\#)$, we have $r(L_x(p))<0$ for some $r \in N_x^+$. This one sees essentially from the fact that $C_x^\#$ is a hyperbolicity cone of $P_x^\#$, since then for every $p\not\in L_x^{-1}(C_x^\#)$ there exists some $v$ on the boundary of $C_x^\#$ such that $DP_x^\#(v)(L_x(p))< 0$, as is shown in the Second Lemma in section \ref{sec_hyperbolicity}. Clearly, the image $DP_x^\#(v)$ of $v$ under the Gauss map $DP_x^\#$ is then a massless covector, and it remains to be shown that it lies inside the positive energy cone $C_x^\#$. Since in an energy-distinguishing spacetime, a null covector is either of positive or of negative energy, it suffices to find a single $y \in C_x^\#$ with $y(DP_x^\#(v))>0$ in order to show that $DP_x^\#(v)$ lies indeed in the positive energy cone $(C_x^\#)^\perp$. But this is easily established from the convexity of $C_x^\#$. For then we certainly find some $y\in C_x^\#$ such that $y+v \in C_x^\#$. But then $y(DP_x^\#(v))=dP_x^\#(v+ s y)/ds|_{s=0} > 0$. In summary, $p \in L_x^{-1}(C_x^\#)$ if and only if there exists an $r\in N_x^+$ with $r(L_x(p))<0$.  

\Section{\revise{Non-physicality of Gambini-Pullin and Myers-Pospelov dispersion relations}}\label{sec_pospelov}
\revise{The developments of this paper are far from academic musings of only remote relevance to physics. Indeed, the identification of bi-hyperbolicity and the energy distingusihing conditions as inevitable properties of dispersion relations provides, once known, a simple algebraic check on the physical consistency of any given dispersion relation. 
How simple indeed it is to apply these conditions has already been shown when we derived that for certain classes of area metric geometries, the general linear electrodynamics formulated on such backgrounds satisfy the physicality conditions. 

In this section we now show that it is equally simple to extract from our results that some rather popular modifications of electrodynamics, namely those of Gambini-Pullin and Myers-Pospelov, indeed possess dispersion relations that reveal the underlying field theory to be non-predictive. In the case of Myers-Pospelov, hyperbolicity (and thus predictivity) can be restored, but unfortunately only at the expense of destroying the energy-distinguishing conditions (and thus a well-defined notion of positive energy). These theories thus do not have the physical interpretation that would be required in order to render observational investigation of bounds on their parameters meaningful. It is obvious that it is both necessary, and above all simple, to subject also any other proposal for modified dispersion relations to the same straightforward tests, and that not to do so is simply negligent from a physical point of view.

\textit{Gambini-Pullin field equations.} Gambini and Pullin \cite{gambini1998} obtained a modified dispersion relation by studying the interaction Hamiltonian for electromagnetic and gravitational fields in a semi-classical approximation motivated by loop quantum gravity.
More precisely, they found the following refined equations for the electromagnetic field
\begin{eqnarray}
\label{fieldequations}
\nabla\times \vec B-\partial_t\vec E+\alpha \nabla^2(\nabla\times\vec B)&=&0\\\nonumber
\nabla\times \vec E+\partial_t \vec B+\alpha \nabla^2(\nabla\times\vec E)&=&0,
\end{eqnarray}
with $\alpha$ being a length scale. In fact, it is easy to see that equations (\ref{fieldequations}) are not well-posed. For if one defines $u^A=(\vec E, \vec B)$, equations (\ref{fieldequations}) become
\begin{equation}
\label{vector_eq}
D_{AB}(\partial)u^B=0,
\end{equation}
with $D_{AB}(\partial)$ a matrix-valued differential operator explicitely given by
\begin{equation}
D_{AB}(\partial)=\left[
\begin{array}{cc}
-\delta_{ik}\partial_t &\epsilon_{ijk}\partial_j +\alpha\, \epsilon_{ijk}\partial_l \partial_l \partial_j\\
\alpha \, \epsilon_{ijk}\partial_j+\epsilon_{ijk}\partial_l \partial_l \partial_j & \delta_{ik}\partial_t
\end{array}\right],
\end{equation}
where in the above expression $\epsilon_{ijk}$ is the standard Levi-Civita symbol and Einstein's summation convention is used. 
The polynomial $\textrm{det } D(iq)$ is easily found to be  
\begin{equation}
\label{polynomial}
P(q)=\textup{det}(D(i q))=-q_0^2 \left( q^2+2 \alpha (\vec q\cdot \vec q)^2- \alpha^2 (\vec q\cdot \vec q)^3  \right)^2,
\end{equation}
with $p^2=p_0^2-\vec p\cdot \vec p$, so that the its principal part 
\begin{equation}
 P(q_0,\vec{q}) = \alpha^4 q_0^2(\vec q\cdot\vec q)^6\,,
\end{equation}
which is not hyperbolic. Even arguing that only lower order of $\alpha$ should be considered, the very same problem remains. Thus, the Gambini-Pullin field equations are not predictive, and the corresponding dispersion relation non-physical. It has been argued \cite{gambini2001} that inadequate quantum states were considered by Gambini and Pullin in the obtainment of equations (\ref{fieldequations}). Urrutia et. al. \cite{alfaro2001} performed a re-examination of Gambini-Pullin calculations, with a more careful motivation for the quantum states considered. However, these result in only slightly different refined equations (neglecting a non-linear term in the magnetic field) for the electromagnetic field, and a very similar analysis as above also shows that again the associated dispersion relation is not hyperbolic. 

\textit{Myers-Pospelov field equations.} Mayers and Pospelov studied dimension 5 operators \cite{myers2003} leading to cubic modified dispersion relations. Specifically, they proposed the following modified equations for the electromagnetic field  
\begin{equation}
 D^\nu_\mu(\partial)A_\nu=0,
\end{equation}
with $D^\nu_\mu(\partial)$ a matrix-valued differential operator explicitely written as
\begin{equation}
\label{myers_op}
 D^\nu_\mu(\partial)=\square \delta^\nu_\alpha+\gamma \,\eta_{\rho \alpha}\,\epsilon^{\rho \sigma \mu \nu} n_\sigma (n\cdot \partial)^2 \partial_\mu,
\end{equation}
where in the above expression, $\gamma$ is the free parameter of the theory, $\eta$ is the standard Lorentzian metric $\eta=\textup{diag}(1,-1,-1,-1)$, and $n$ is a time-like covector with respect to $\eta$, i.e. $\eta(n,n)>0$, which breakes Lorentz invariance. For the operator (\ref{myers_op}) one finds
 \begin{equation}
 \label{polynomial_Pospelov}
 \textup{det}(D(i q))=(q^2)^2 \left[(q^2)^2-\gamma ^2 (n\cdot q)^4 \left(n^\alpha n^\beta -\eta^{\alpha\beta} n^2\right)q_\alpha q_\beta\right].
\end{equation}
From the above expression we read off the principal part 
\begin{equation}
 P(q) = \gamma ^2 (p^2)^2 (n\cdot p)^4 \left(n^\alpha n^\beta -\eta^{\alpha\beta} n^2\right)p_\alpha p_\beta\,,
\end{equation}
which for $\eta(n,n)>0$ is not hyperbolic. This is so, because the matrix $n^\alpha n^\beta -\eta^{\alpha\beta} n^2 $, under the assumption $\eta(n,n)>0$, is positive semi-definite, which implies that one of the factors of the principal part of $P(p)$, namely $\left(n^\alpha n^\beta -\eta^{\alpha\beta} n^2\right)p_\alpha p_\beta$ is not hyperbolic. Hence, the Myers-Pospelov field equations are non-predictive. Furthermore, even if one were to choose $n$ such that $\eta(n,n)\leq 0$, one would still have a null-plane due to the term $n\cdot p$, which we saw at the end of section \ref{sec_bihyperbolicity} to obstruct the energy distinguishing property, and thus to lead to a non-physical dispersion relation. Precisely the same argument also rules out the scalar and fermionic modified field equations presented in \cite{myers2003}.

In conclusion, the field equations found by Gambini, Pullin and Urrutia in the framewok of loop quantum gravity, as well as the field equations found by Myers and Pospelov in the framework of effective field theory do not lead to physical dispersion relations. More precisely, there is no spacetime hypersurface $\Sigma$ on which initial data for the electromagnetic field  could be given so that its values on a later hypersurface would be uniquely prescribed. Hence, phenomenological conclusions, such as the identification of bounds, based on these modified dispersion relations are unfortunately not conclusive. }

\Section{Conclusions}\label{sec_conclusions}
Our investigation of viable dispersion relations in classical physics yielded far more than the identification of the three restrictive conditions we aimed at. Indeed, while deriving that a dispersion relation must be encoded in a cotangent bundle function $P$ that (i) defines a reduced homogeneous polynomial in each cotangent space (ii) is hyperbolic and has a dual tangent bundle function that is also hyperbolic and (iii) is energy-distinguishing, we collected a number of further important results along the way.

First of all, the algebraic properties to be satisfied by a dispersion relation immediately restrict also the possible fundamental spacetime geometries from which the dispersion relation derives. While the connection between the dispersion relation (encoded in a cotangent bundle function $P$) and the fundamental spacetime geometry to which fields couple (encoded in some tensor $G$) played an explicit r\^ole only in sections \ref{sec_covdisprelmassless} and \ref{sec_hyperbolicity}, all the restrictions we derived in this paper for the cotangent bundle function $P$ automatically \new{translate into} restrictions on the geometry $G$. Making this transparent has been one purpose of the illustrations throughout, where we saw how a fundamental geometry given by a symmetric second rank tensor, for instance, is automatically restricted to be a Lorentzian geometry (due to the requirement of bi-hyperbolicity of the induced dispersion relation), and then automatically energy-distinguishing. Thus we recovered, from first principles, the familiar result that Lorentzian manifolds provide a consistent fundamental spacetime structure. The crucial point, of course, is that the very same principles and resulting conditions apply for any other tensorial geometry. \revise{We illustrated that while these physicality conditions do hold for certain classes of area metric geometry, they do not for two very popular proposals for field theories with modified dispersion relations.}

Second, we obtained a complete theory of observers and point particles. One remarkable observation here was that massless and massive particles are governed by very different geometries on the tangent bundle. While massive particles are governed by a generically non-polynomial Finsler function $P^*$, it is a in each fibre homogeneously polynomial tangent bundle function $P^\#$ that governs the motion of massless particles. But both tangent space structures derive from the same cotangent bundle structure given by $P$. The distinction is generated due to fundamentally different duality maps between cotangent and tangent spaces, depending on whether one is dealing with massive or massless momenta. While in the former case the duality is given by a Legendre map, it is a projective Gauss map in the latter case. In both cases, the duality maps are generically non-linear, which replaces the ubiquitous linear algebra in metric geometry by the need for some elementary convex analysis and real algebraic geometry in the general case.   
  
Third, we found that even for modified dispersion relations, key properties of the Lorentzian kinematics still hold. One result is that observers can always be thought of as massive. Another one that the decay of a massless particle into massive ones is kinematically impossible. A third kinematical issue, namely the kinematic exclusion of processes where a massive particle radiates off energy in terms of massless particles, only holds below a certain energy threshold. The energy threshold, in turn, is encoded in a subcone of the positive energy massive momentum cone and thus presents a fully covariant notion. We find that while observers automatically respect that energy threshold, there are massive particle momenta exceeding it. But then a Cerenkov-type process, by which the massive particle radiates off energy in terms of a massless particle, is kinematically possible even in vacuo. Proving these assertions, which amount to a covariant geometric mechanism for an energy cut-off, required use of the entire machinery we developed for general dispersion relations. Of course, one cannot make any statement about the average decay time without studying the quantum theory of matter on such generalized geometries. 

Fourth, we saw that a non-covariant representation of dispersion relations, where the energy of a point particle seen by a particular observer is given as a function of its spatial momentum, is not even meaningful all  by itself. The trouble is that the decomposition of a spacetime momentum into its energy and purely spatial momentum requires a temporal-spatial split, which in \new{turn} depends on the particular observer, \new{by way of} the cotangent bundle function $P$. This means that the kinematical objects appearing in a non-covariant dispersion relation can only be given meaning by first obtaining the covariant formulation. In practice therefore, it is hard to start with a non-covariant formulation. Even worse, there may well be no covariant formulation at all from which a given non-covariant relation could be derived, in which case the latter is revealed to be in fact meaningless. It is therefore sensible, both conceptually and physically, to start from the covariant formulation, as we have done throughout this article. Even more so, since the crucial algebraic properties we identified for the cotangent bundle function are deeply hidden in a non-covariant formulation.   

Fifth, even the notion of freely falling and non-rotating frames hinges on the covariant dispersion relation at hand, and is provided by a generically non-linear connection on the tangent bundle. Thus the dispersion relation is also seen to have an effect on the interpretation of spacetime quantities (such as for instance the electromagnetic field strength two-form) in terms of quantities that are actually measurable in some laboratory and related to the covariant quantity through an observer frame (such as electric and magnetic fields). It is both remarkable, and important to realize, that the choice of a dispersion relation has such far-reaching implications. At the same time, our algebraic restrictions on the cotangent bundle function $P$, originally required for other reasons, single-handedly ensure that all required kinematical notions exist \new{can be constructed}. 

Sixth, the problem of finding a pseudo-Finslerian analogue of Lorentzian geometry is solved. The twist required by our findings is that the geometry is established by a single function $P$ on the cotangent bundle, rather than the tangent bundle. Moreover, one can and must restrict attention to functions which are bi-hyperbolic and energy-distinguishing reduced homogeneous polynomials in each cotangent fibre. While we saw that all of these properties, and their interplay, are important for the geometry to provide a viable spacetime structure, it is bi-hyperbolicity in particular which generalizes the Lorentzian character of metrics to the much more general geometries studied in this work. Indeed, the same physical principle that selects Lorentzian signature amongst all possible metric geometries, namely the conditions of having well-posed matter field equations and a well-defined notion of positive particle energy, directly led to the above general conditions that cover also non-metric geometries. So while on the cotangent bundle, there is one single geometric structure over which we have excellent mathematical control, there are two very different structures induced by it on the tangent bundle, the duals $P^*$ and $P^\#$. \new{} The intuition behind many studies concerning a Finslerian generalization of general relativity has been correct in allowing for a very generic (non-polynomial) structure $P^*$ on the tangent spaces in order to describe massive point particles. However, we now know that attempts to describe massless point particles with the same structure must fail in general: The massless particle motion governed by the polynomial structure $P^\#$ will generically not coincide with the non-polynomial $P^*$ governing massive particle dynamics. So not even the most general Finslerian or Lagrangian geometry on the tangent bundle could possibly serve as a viable spacetime geometry in any other than the most special circumstances (such as the geometry actually being metric). Starting from the cotangent bundle, instead, one has the here developed theory available.\\

With the results obtained in this work, we are now well equipped to \new{address} two key questions, which arise whenever a modified dispersion relation (or, equivalently, a modified fundamental spacetime geometry) \new{is} considered. 

The first issue is that one needs well-posed dynamical equations that can replace the Einstein equations. Remarkably, this apparently physical question is in fact a predominantly mathematical one: Dynamics for a particular spacetime geometry can be obtained from studying minimal representations of the deformation algebra of hypersurfaces in that very same geometry, so that the field equations are well-posed by construction. The geometric key ingredients are to have available, first, a Legendre map associating the normal covectors of suitable initial data surfaces with their vector duals, and second, a notion of proper time that \new{provides a physically distinguished} normalization of those normal vectors.
 For the Lorentzian case, this program has been carried out in seminal work by Hojman, Kucha\u r and Teitelboim \cite{hojman1976}, and leads to the Einstein-Hilbert action with an undetermined cosmological constant as the unique such dynamics.
 The geometries identified in the present work are just the ones where these tools are indeed still available and thus the corresponding hypersurface deformation algebras can be derived straightforwardly. Finding respresentations in terms of the geometric variables directly leads to the corresponding non-metric gravity theories whose dynamics are well-posed by construction. In other words, with the tools developed in this paper, the physical art of constructing an alternative gravity theory for a given geometric structure has been reduced to a well-defined mathematical question in representation theory. \new{}     

On the other hand, one needs to understand the quantum theory of particles and fields. Equipped with the technical machinery we developed, one may now indeed start from the massive or massless point particle action and perform a first quantization, thereby obtaining field equations. Subtleties, which one can gloss over in the case of a Lorentzian background geometry without harm, now need to be taken into account. For instance, the restriction that momenta must lie within the hyperbolicity cone for massive particles, or on the cone of null covectors for massless particles, becomes of crucial technical importance in the non-metric case. In particular, this has important repercussions for a second quantization of the field equations one obtained from the first quantization. Again, as in the question of identifying appropriate gravitational dynamics, the actual execution of the quantization may present, depending on the chosen geometry, \new{a hard, albeit now well-defined,} mathematical problem.

\section*{Acknowledgments}
FPS thanks Cedric Deffayet, Kristina Giesel and Christof Witte for insightful remarks and discussions. The work of SR has been supported by a doctoral research scholarship of the German Academic Exchange Service DAAD and that of DR by the International Max Planck Research School for Geometric Analysis, Gravitation and String Theory.

%\newpage
\bibliographystyle{ieeetr} %plainnat
\bibliography{dispersion}

\end{document}